# Impact of mode completeness on the accuracy of the coupling theory of quasinormal modes: a strict numerical demonstration

Can Tao,[1,2,3] Junda Zhu[4], and Haitao Liu[1,2,*]

[1] *Institute of Modern Optics, College of Electronic Information and Optical Engineering, Nankai University, Tianjin, 300350, China*

[2] *National Key Laboratory of Semiconductor Laser, Tianjin, 300350, China*

[3] *The MOE Key Laboratory of Weak Light Nonlinear Photonics, TEDA Institute of Applied Physics and School of Physics, Nankai University, Tianjin 300457, China*

[4] *College of Physics and Materials Science, Tianjin Normal University, Tianjin 300387, China*

[*] Corresponding author: liuht@nankai.edu.cn

The coupling theory of quasinormal modes (QNMs) for a coupled system of generally lossy and dispersive optical nanoresonators has been established in a rigorous manner based on the first principle of Maxwell's equations [Phys. Rev. B **102**, 045430 (2020)], and can achieve superior computational efficiency and physical intuitiveness compared with full-wave numerical methods if a small set of modes can achieve a high accuracy. The QNMs suffer from an exponential divergence of far field and can form a complete basis inside but not outside the resonator. In the QNM coupling theory (QCT), it is required that the QNMs of each resonator form a complete basis in expanding the scattered field both inside and outside the resonator, which can be achieved by using the regularized QNMs (RQNMs). However, a strict numerical demonstration of the impact of the mode completeness of RQNMs on the accuracy of QCT by using a virtually complete basis of RQNMs is still absent. In this paper, we will provide such a numerical demonstration along with an improvement of the QCT and some theoretical demonstrations on a rigorous incorporation of the RQNMs into the QCT. The RQNMs are obtained by introducing an equivalent surface current (ESC) encircling the resonator (called ESC-RQNMs) or the perfectly matched layer (PML) surrounding the whole computational domain (called PML-RQNMs). The numerical example is selected as two one-dimensional (1D) resonators of slabs in the extreme coupling case of direct contact, for which a virtually complete basis of RQNMs can be solved either analytically (for ESC-RQNMs) or numerically (for PML-RQNMs). The results show that by using a virtually complete basis of RQNMs, the QCT can achieve a high accuracy in predicting both the source-free eigenmodes and the source-excited scattered field of the coupled system, which is not true if using the incomplete basis of not-regularized QNMs (i.e., physical QNMs).

## I. INTRODUCTION

The systems composed of multiple optical resonators are widely used for constructing nanophotonic devices [1-3]. The optical performances of such coupled systems are largely dependent on the coupling effects among resonators, and can be modeled by the coupling theory of eigenmodes [4-8] in a computationally-efficient and physically-intuitive way compared with the full-wave numerical methods [9]. The eigenmode of non-conservative (non-Hermitian) optical system can be rigorously defined as quasinormal mode (QNM) [5,10-15], which is the eigensolution of source-free Maxwell's equations at a complex eigenfrequency and satisfies outgoing-wave condition at infinity. In recent years, several coupling theories of QNMs have been developed [4-6,16-18] for studying non-Hermitian coupled systems, and are applied for instance in predicting the rich spectral responses at exceptional points [16], tailoring the frequency spectrum of Purcell factor [6] and realizing zero coupling between nearby cavities [18]. Among them, the QNM coupling theory (QCT) proposed in Ref. [6] by some of the authors of this paper is generally applicable to systems with an arbitrary frequency dispersion of permittivity and in a possibly inhomogeneous background medium with or without the presence of excitation source. The theory is derived in a rigorous manner from the first principles of Maxwell's equations under the only assumption that the QNMs of each resonator in the coupled system can form a complete basis in expanding the scattered field of each resonator. Due to the difficulty in solving all the QNMs of each resonator with a generally three-dimensional (3D) geometry, the numerical tests of the QCT are performed only in an approximate manner using few dominant QNMs [6], and a strict numerical test of the accuracy of the QCT when using a virtually complete basis of QNMs is still absent, which is crucial for justifying the theoretically expected rigorousness of the QCT and the resultant ability for providing quantitative predictions.

As mentioned above, the completeness of QNM basis in expanding source-excited electromagnetic field is the key to ensure the rigorousness and the resultant accuracy of the QCT. For resonators of analytically-solvable simple geometries such as one-dimensional (1D) slab Fabry-Perot resonators [10] or spherical Mie-resonators [19] in a homogeneous background medium, it has been proven that the analytically-solvable physical QNMs form a complete basis in the interior (defined by the discontinuities of permittivity) but not in the exterior of the resonator. Here the physical QNMs (or called not-regularized QNMs) are defined as the eigensolutions of the exact Maxwell's equations without discretization. For resonators of general geometries in a uniform background medium, numerical results show that the physical QNMs form a complete basis inside but not outside the resonator [15], which however still lacks a rigorous theoretical proof at present.

On the other hand, it is demonstrated that physical QNMs along with the virtual gap modes generated by the resonant-state expansion [20] can form a complete basis in a region larger than the resonator but at the expense of solving more modes and resultantly larger computational amount [17]. The problem caused by the incompleteness of QNMs can also be handled by adding a continuum of modes, building on which, a rigorous QCT was recently developed [17] to predict the eigenmodes of coupled resonators.

To achieve the completeness of modes in reproducing the field both inside and outside the resonator, the perfectly matched layers (PMLs) are proposed to regularize and normalize QNMs, so that the outgoing far field of QNMs is transformed from exponential divergence in physical space to exponential decay in the PMLs [12]. Afterwards, a number of numerical results indicate that such PML-regularized QNMs (PML-RQNMs) [15] can be complete to ensure an accurate expansion of scattered field both inside and outside a resonator, and this numerical completeness of finite number of PML-RQNMs can be understood from the viewpoint of PMLs-discretized Maxwell's equations with finite degrees of freedom [14,21,22].

Another approach of regularizing QNMs is to introduce a regularized field outside the resonator [23-26], which is the radiation field of an equivalent current at the real excitation frequency and thus is free of the far-field divergence of QNMs. Compared with the equivalent volumetric current located within the domain of the resonator [23,24], the equivalent surface current (ESC) encircling the resonator requires less computation [25,26] for regularizing QNMs (called ESC-RQNMs). Note that the ESC-RQNM is conceptually different from the constant flux state [27,28]. A theoretical demonstration on the completeness of the ESC-RQNMs in expanding the scattered field both inside and outside a resonator is provided for resonators in a homogeneous background medium [25] or on a flat-interface substrate [26], but is still lacking for the general case of resonators in an arbitrary inhomogeneous background medium.

The validity of the QCT proposed in Ref. [6] using few modes of the PML-RQNMs [6,29] or ESC-RQNMs [25,26] has been tested numerically. However, a strict numerical verification of the QCT requires the use of a virtually complete basis of RQNMs. For a theory [5] methodologically equivalent to QCT using ESC-RQNMs [25,26] (although the concept of ESC-RQNM is not explicitly proposed in the theory [5]), convergence analysis by using progressively increasing number of modes has been implemented in predicting the eigenmodes of two coupled slabs [5], but is absent in predicting the scattered field. For the QCTs established in Refs. [4] and [18], the convergence in predicting the eigenfrequency [4,18] or the scattered field [18] was numerically studied by using partial PML-RQNMs.

In this paper, we report a strict numerical demonstration of the accuracy of the QCT developed in Ref. [6] in predicting both source-free eigenmodes and source-excited scattered field of a coupled system when using a virtually complete basis of PML-RQNMs or ESC-RQNMs, and also propose an improvement of the QCT and some theoretical demonstrations on a rigorous incorporation of the RQNMs into the QCT, for instance, a theoretical demonstration on the completeness of the ESC-RQNMs for the general case that the resonator can be located in an arbitrary inhomogeneous background medium. The numerical example is selected to be two one-dimensional (1D) resonators of slabs in an extreme coupling case of direct contact, which has the merit that a virtually complete basis of RQNMs can be solved either analytically (for ESC-RQNMs) or numerically (for PML-RQNMs). The results show that by using the virtually complete basis of RQNMs, the QCT can achieve a high accuracy in predicting both the eigenmodes and the scattered field, which however is not true if using the incomplete basis of not-regularized QNMs (i.e., physical QNMs).

This paper is organized as follows. In Sec. II, we present a brief review and the improvement of the QCT, along with the completeness of ESC-RQNMs and PML-RQNMs and their incorporation into the QCT. In Sec. III, the numerical demonstration of the accuracy of the QCT when using a virtually complete basis of RQNMs is presented. Conclusions are summarized in Sec. IV.

## II. METHODS

### A. A brief review and improvement of the QCT

In the previous work [6], the QCT applicable to generally lossy and dispersive optical resonators is built up in a rigorous manner based on the first principles of Maxwell's equations. The theory can comprehensively predict both the source-free QNMs and source-excited scattered field of the coupled system, which are expressed in terms of the QNMs of each single resonator in the system. As a preparation of this work, in the following we will provide a brief review along with some improvement of the QCT.

For a system composed of $P$ scatterers of resonators, the permittivity tensor $\boldsymbol{\varepsilon}(\mathbf{r},\omega)$ [dependent on spatial coordinates $\mathbf{r}=(x,y,z)$ and angular frequency $\omega$] can be expressed as $\boldsymbol{\varepsilon}(\mathbf{r},\omega)=\boldsymbol{\varepsilon}_{\mathrm{b}}(\mathbf{r},\omega)+\sum_{p=1}^{P}\Delta\boldsymbol{\varepsilon}_p(\mathbf{r},\omega)$, where $\boldsymbol{\varepsilon}_{\mathrm{b}}$ is the permittivity of the background medium, and $\Delta\boldsymbol{\varepsilon}_p=\boldsymbol{\varepsilon}_p-\boldsymbol{\varepsilon}_{\mathrm{b}}$ is the permittivity change of the $p$th scatterer with $\boldsymbol{\varepsilon}_p$ being the permittivity with the presence of only the $p$th scatterer, so $\Delta\boldsymbol{\varepsilon}_p$ is zero outside the $p$th scatterer. The permeability tensor $\boldsymbol{\mu}(\mathbf{r},\omega)$ can be expressed similarly. For nonmagnetic medium considered here, there is $\boldsymbol{\mu}(\mathbf{r},\omega)=\mu_0$ being the permeability of vacuum. The scattered field $\boldsymbol{\Psi}^{\mathrm{sca}}(\mathbf{r},\omega)=[\mathbf{E}^{\mathrm{sca}}, \mathbf{H}^{\mathrm{sca}}]^{\mathrm{T}}$ defined as the total field $\boldsymbol{\Psi}(\mathbf{r},\omega)$ (with the presence of all the $P$ scatterers) minus the background field $\boldsymbol{\Psi}^{\mathrm{b}}(\mathbf{r},\omega)=[\mathbf{E}^{\mathrm{b}}, \mathbf{H}^{\mathrm{b}}]^{\mathrm{T}}$ (without any scatterer) satisfies the Maxwell's equations,

$$\begin{aligned}\nabla\times\mathbf{E}^{\mathrm{sca}}&=i\omega\mu_0\mathbf{H}^{\mathrm{sca}},\\ \nabla\times\mathbf{H}^{\mathrm{sca}}&=-i\omega\boldsymbol{\varepsilon}\cdot\mathbf{E}^{\mathrm{sca}}-i\omega\sum_{p=1}^{P}\Delta\boldsymbol{\varepsilon}_p\cdot\mathbf{E}^{\mathrm{b}},\end{aligned}\tag{1}$$

and satisfies the outgoing-wave condition at infinity. A temporal dependence $\exp(-i\omega t)$ of the electromagnetic field is assumed in this paper. It can be proven that Eq. (1) has a solution in the following form [see Eqs. (S5) and (S6) in Ref. [6] for the proof],

$$\boldsymbol{\Psi}^{\mathrm{sca}}(\mathbf{r},\omega)=\sum_{p=1}^{P}\boldsymbol{\Psi}_p^{\mathrm{sca}}(\mathbf{r},\omega),\tag{2}$$

where $\boldsymbol{\Psi}_p^{\mathrm{sca}}=[\mathbf{E}_p^{\mathrm{sca}},\mathbf{H}_p^{\mathrm{sca}}]^{\mathrm{T}}$ satisfies the Maxwell's equations,

$$\begin{aligned}\nabla\times\mathbf{E}_p^{\mathrm{sca}}&=i\omega\mu_0\mathbf{H}_p^{\mathrm{sca}},\\ \nabla\times\mathbf{H}_p^{\mathrm{sca}}&=-i\omega\boldsymbol{\varepsilon}_p\cdot\mathbf{E}_p^{\mathrm{sca}}-i\omega\Delta\boldsymbol{\varepsilon}_p\cdot\left(\sum_{q\neq p}\mathbf{E}_q^{\mathrm{sca}}+\mathbf{E}^{\mathrm{b}}\right),\end{aligned}\tag{3}$$

where $\boldsymbol{\varepsilon}_p=\boldsymbol{\varepsilon}_{\mathrm{b}}+\Delta\boldsymbol{\varepsilon}_p$ is the permittivity tensor with the presence of only the $p$th scatterer in the background medium. Equation (3) shows that the $\boldsymbol{\Psi}_p^{\mathrm{sca}}$ is the scattered field of the $p$th scatterer excited by $\boldsymbol{\Psi}^{\mathrm{b}}$ and the scattered fields $\sum_{q\neq p}\boldsymbol{\Psi}_q^{\mathrm{sca}}$ of the other scatterers, which is simply a rigorous formulation of the intuitive multiple-scattering picture.

The $m$th-order QNM $\boldsymbol{\Psi}_{p,m}(\mathbf{r})=[\mathbf{E}_{p,m}, \mathbf{H}_{p,m}]^{\mathrm{T}}$ of the $p$th resonator is the eigensolution of the source-free Maxwell's equations,

$$\begin{aligned}\nabla\times\mathbf{E}_{p,m}&=i\omega_{p,m}\mu_0\mathbf{H}_{p,m},\\ \nabla\times\mathbf{H}_{p,m}&=-i\omega_{p,m}\boldsymbol{\varepsilon}_p(\mathbf{r},\omega_{p,m})\cdot\mathbf{E}_{p,m},\end{aligned}\tag{4}$$

and satisfies the outgoing-wave condition at infinity, where $\omega_{p,m}$ is the complex eigenfrequency. According to the QNM expansion formalism, the scattered field $\boldsymbol{\Psi}_p^{\mathrm{sca}}$ of the $p$th resonator can be expressed as a linear superposition of the $\boldsymbol{\Psi}_{p,m}$,

$$\boldsymbol{\Psi}_p^{\mathrm{sca}}(\mathbf{r},\omega)=\sum_{m=1}^{M_p}\alpha_{p,m}(\omega)\boldsymbol{\Psi}_{p,m}(\mathbf{r}),\tag{5}$$

where the QNM-expansion coefficient $\alpha_{p,m}$ can be expressed in terms of the excitation source and QNMs as [6,12,30],

$$\begin{aligned}\alpha_{p,m}(\omega)=\frac{-\omega}{(\omega-\omega_{p,m})F_{p,m}}&\iiint_{V_p}\mathbf{E}_{p,m}(\mathbf{r})\cdot\Delta\boldsymbol{\varepsilon}_p(\mathbf{r},\omega)\\ &\cdot\left[\sum_{q\neq p}\mathbf{E}_q^{\mathrm{sca}}(\mathbf{r},\omega)+\mathbf{E}^{\mathrm{b}}(\mathbf{r},\omega)\right]d^3\mathbf{r},\end{aligned}\tag{6}$$

where the method for calculating the pseudoenergy $F_{p,m}$ is explained in Sec. S1 of Supplemental Material (SM) [31], and $V_p$ denotes the region of the $p$th resonator out of which there is $\Delta\boldsymbol{\varepsilon}_p=\mathbf{0}$.

Substituting Eq. (5) into Eq. (6) to replace the $\mathbf{E}_q^{\mathrm{sca}}$ by $\sum_{n=1}^{M_q}\alpha_{q,n}(\omega)\mathbf{E}_{q,n}(\mathbf{r})$, one can obtain the QNM-coupling equations,

$$\beta_{p,m}(\omega)=(\omega-\omega_{p,m})\alpha_{p,m}(\omega)-\sum_{q\neq p}\sum_{n=1}^{M_q}\kappa_{(p,m),(q,n)}(\omega)\alpha_{q,n}(\omega),\tag{7}$$

where the QNM-coupling coefficient $\kappa_{(p,m),(q,n)}$ is defined as

$$\kappa_{(p,m),(q,n)}(\omega)=-\frac{\omega}{F_{p,m}}\iiint_{V_p}\mathbf{E}_{p,m}(\mathbf{r})\cdot\Delta\boldsymbol{\varepsilon}_p(\mathbf{r},\omega)\cdot\mathbf{E}_{q,n}(\mathbf{r})d^3\mathbf{r},\tag{8}$$

which is the coefficient of the $m$th-order QNM of the $p$th resonator excited by the $n$th-order QNM of the $q$th resonator. The QNM-excitation coefficient $\beta_{p,m}$ is defined as

$$\beta_{p,m}(\omega)=-\frac{\omega}{F_{p,m}}\iiint_{V_p}\mathbf{E}_{p,m}(\mathbf{r})\cdot\Delta\boldsymbol{\varepsilon}_p(\mathbf{r},\omega)\cdot\mathbf{E}^{\mathrm{b}}(\mathbf{r},\omega)d^3\mathbf{r},\tag{9}$$

which is the coefficient of the $m$th-order QNM of the $p$th resonator excited by the background field $\boldsymbol{\Psi}^{\mathrm{b}}$. Equation (7) can be rewritten as a concise matrix form,

$$\mathbf{A}(\omega)\mathbf{a}(\omega)=\mathbf{b}(\omega),\tag{10}$$

where the $(p, m)$th element of the column vector $\mathbf{a}(\omega)$ is $\alpha_{p,m}$, and the $(p, m)$th element of the column vector $\mathbf{b}(\omega)$ is $\beta_{p,m}$. The element of the $(p, m)$th row and $(q, n)$th column of the matrix $\mathbf{A}(\omega)$ is $-\kappa_{(p,m),(q,n)}$ for $p\neq q$, and is $\omega-\delta_{m,n}\omega_{p,m}$ for $p=q$, where $\delta_{m,n}$ is the Kronecker delta. The QNM-expansion coefficients $\mathbf{a}(\omega)$ can be obtained by solving the set of linear equations (10), and then the scattered field of the coupled system can be obtained as,

$$\boldsymbol{\Psi}^{\mathrm{sca}}(\mathbf{r},\omega)=\boldsymbol{\Phi}(\mathbf{r})^{\mathrm{T}}\mathbf{a}(\omega)=\boldsymbol{\Phi}(\mathbf{r})^{\mathrm{T}}\mathbf{A}(\omega)^{-1}\mathbf{b}(\omega),\tag{11}$$

where $\boldsymbol{\Phi}(\mathbf{r})^{\mathrm{T}}$ is a row vector with its $(p, m)$th element being $\boldsymbol{\Psi}_{p,m}$. A superposition of the scattered field and the background field then gives the total field, i.e., $\boldsymbol{\Psi}=\boldsymbol{\Psi}^{\mathrm{sca}}+\boldsymbol{\Psi}^{\mathrm{b}}$.

Removing the excitation term $\mathbf{b}(\omega)$, Eq. (10) becomes,

$$\mathbf{A}(\omega)\mathbf{a}(\omega)=\mathbf{0}.\tag{12}$$

Then the source-free QNMs of the coupled system can be obtained by solving the nontrivial solutions of Eq. (12), whose existence requires

$$\det\left[\mathbf{A}(\omega)\right]=0. \tag{13}$$

The solutions of Eq. (13) are just the complex eigenfrequencies $\omega=\tilde{\omega}_r$ of the coupled system. Substituting the solved eigenfrequency $\omega=\tilde{\omega}_r$ into Eq. (12), one can then solve its nontrivial solution $\mathbf{a}=\tilde{\mathbf{a}}_r$, and finally obtain the QNM field $\tilde{\mathbf{\Psi}}_r(\mathbf{r})$ of coupled system through $\tilde{\mathbf{\Psi}}_r(\mathbf{r})=\mathbf{\Phi}(\mathbf{r})^{\mathrm{T}}\tilde{\mathbf{a}}_r$. For an isotropic medium with a permittivity of frequency dispersion described by an $N$-pole Drude-Lorentz model [36], Eq. (12) can reduce to a polynomial matrix eigenvalue problem [6,37], for which well-developed algorithms [38] are available to solve all the eigen solutions.

To expand the scattered field $\mathbf{\Psi}^{\mathrm{sca}}$ of the coupled system upon the basis of $\tilde{\mathbf{\Psi}}_r$, so as to explicitly show the impact of $\tilde{\mathbf{\Psi}}_r$ on the resonance properties of $\mathbf{\Psi}^{\mathrm{sca}}$, Eq. (11) can be rewritten as,

$$\mathbf{\Psi}^{\mathrm{sca}}(\mathbf{r},\omega)=\frac{\mathbf{\Phi}(\mathbf{r})^{\mathrm{T}}\mathbf{A}(\omega)^{*}\mathbf{b}(\omega)}{\det[\mathbf{A}(\omega)]}, \tag{14}$$

where $\mathbf{A}(\omega)^{-1}=\mathbf{A}(\omega)^{*}/\det[\mathbf{A}(\omega)]$ is used, with $\mathbf{A}(\omega)^{*}$ denoting the adjugate matrix of $\mathbf{A}(\omega)$. Equations (13) and (14) show that the eigenfrequency $\tilde{\omega}_r$ is the complex pole of $\mathbf{\Psi}^{\mathrm{sca}}$. Then applying the complex-pole expansion theorem (Mittag-Leffler theorem) [39], one can obtain [6],

$$\mathbf{\Psi}^{\mathrm{sca}}(\mathbf{r},\omega)=\mathbf{\Psi}^{\mathrm{sca}}(\mathbf{r},0)+\sum_r\frac{\omega/\tilde{\omega}_r}{\omega-\tilde{\omega}_r}\mathrm{Res}[\mathbf{\Psi}^{\mathrm{sca}}(\mathbf{r},\omega),\tilde{\omega}_r], \tag{15}$$

where

$$\mathrm{Res}[\mathbf{\Psi}^{\mathrm{sca}}(\mathbf{r},\omega),\tilde{\omega}_r]=\lim_{\omega\to\tilde{\omega}_r}(\omega-\tilde{\omega}_r)\mathbf{\Psi}^{\mathrm{sca}}(\mathbf{r},\omega)$$

represents the residue of $\mathbf{\Psi}^{\mathrm{sca}}$ at the complex pole $\tilde{\omega}_r$. Equation (15) holds under the condition that $\mathbf{\Psi}^{\mathrm{sca}}(\mathbf{r},\omega)$ is a meromorphic function of $\omega$ and is bounded as $\omega\to\infty$ in the complex plane, and that $\tilde{\omega}_r$ (satisfying $\tilde{\omega}_r\neq 0$) is the first-order pole of $\mathbf{\Psi}^{\mathrm{sca}}$. After calculating the residue in Eq. (15) and neglecting $\mathbf{\Psi}^{\mathrm{sca}}(\mathbf{r},0)$, one can finally obtain the expansion of the scattered field $\mathbf{\Psi}^{\mathrm{sca}}$ upon the basis of the QNM field $\tilde{\mathbf{\Psi}}_r$ of the coupled system [6],

$$\mathbf{\Psi}^{\mathrm{sca}}(\mathbf{r},\omega)=\sum_r\frac{\omega/\tilde{\omega}_r}{\omega-\tilde{\omega}_r}\frac{\mathbf{h}_r^{\mathrm{T}}\mathbf{b}(\tilde{\omega}_r)}{\partial\det[\mathbf{A}(\omega)]/\partial\omega|_{\omega=\tilde{\omega}_r}}\tilde{\mathbf{\Psi}}_r(\mathbf{r}), \tag{16}$$

where $\mathbf{h}_r^{\mathrm{T}}$ is a row vector determined by $\mathbf{A}(\tilde{\omega}_r)^{*}=\tilde{\mathbf{a}}_r\mathbf{h}_r^{\mathrm{T}}$.

For the QNM expansion of Eq. (16), the excitation source of background field $\mathbf{\Psi}^{\mathrm{b}}(\mathbf{r},\omega)$ should fulfill the following requirements:

i) As $\omega\to\infty$ in the complex plane, $\mathbf{\Psi}^{\mathrm{b}}(\mathbf{r},\omega)$ is bounded so that $\mathbf{\Psi}^{\mathrm{sca}}(\mathbf{r},\omega)$ is bounded as required by Eq. (15);

ii) $\mathbf{\Psi}^{\mathrm{b}}(\mathbf{r},\omega)$ has no pole (i.e., analytic) in the complex plane of $\omega$. Otherwise, the pole of $\mathbf{\Psi}^{\mathrm{b}}(\mathbf{r},\omega)$ should be included into the $\tilde{\omega}_r$ in Eq. (15), which then violates the requirement by Eq. (16) that the $\tilde{\omega}_r$ only includes the zeros of $\det[\mathbf{A}(\omega)]$.

The above requirements i) and ii) lead to a consequence that $\mathbf{\Psi}^{\mathrm{b}}(\mathbf{r},\omega)$ should be independent of $\omega$, which is a severe limitation. For example, the common background field of a plane wave propagating in free space (with a refractive index $n_1$) along $z$-direction, which is adopted in the numerical example of this paper, has a frequency dependence of $\exp(ik_0n_1z)$ with $k_0=\omega/c$ ($c$ being the light speed in the vacuum). To remove the above limitation, in the following, we will apply the complex-pole expansion theorem to the $\mathbf{A}(\omega)^{-1}=\mathbf{A}(\omega)^{*}/\det[\mathbf{A}(\omega)]$ instead of $\mathbf{\Psi}^{\mathrm{sca}}$ in Eq. (14), i.e.,

$$\mathbf{A}(\omega)^{-1}=\mathbf{A}(\infty)^{-1}+\sum_r\frac{1}{\omega-\tilde{\omega}_r}\mathrm{Res}[\mathbf{A}(\omega)^{-1},\tilde{\omega}_r], \tag{17}$$

where $\tilde{\omega}_r$ is the complex pole of $\mathbf{A}(\omega)^{-1}$ as shown in Eq. (13), and it is assumed that $\mathbf{A}(\omega)^{-1}$ [i.e., each element of $\mathbf{A}(\omega)^{-1}$] is a meromorphic function of $\omega$ and is bounded as $\omega\to\infty$ in the complex plane, and that $\tilde{\omega}_r$ is the first-order pole of $\mathbf{A}(\omega)^{-1}$. Note that different from Eq. (15) which requires $\tilde{\omega}_r\neq 0$, Eq. (17) allows $\tilde{\omega}_r=0$ which corresponds to the static modes [14]. Then similarly to the calculations in Ref. [6], the residue in Eq. (17) can be calculated as,

$$\begin{aligned}\mathrm{Res}[\mathbf{A}(\omega)^{-1},\tilde{\omega}_r]&=\lim_{\omega\to\tilde{\omega}_r}(\omega-\tilde{\omega}_r)\mathbf{A}(\omega)^{-1}\\&=\lim_{\omega\to\tilde{\omega}_r}(\omega-\tilde{\omega}_r)\frac{\mathbf{A}(\omega)^{*}}{\det[\mathbf{A}(\omega)]}\\&=\frac{\mathbf{A}(\tilde{\omega}_r)^{*}}{\partial\det[\mathbf{A}(\omega)]/\partial\omega|_{\omega=\tilde{\omega}_r}}\\&=\frac{\tilde{\mathbf{a}}_r\mathbf{h}_r^{\mathrm{T}}}{\partial\det[\mathbf{A}(\omega)]/\partial\omega|_{\omega=\tilde{\omega}_r}}.\end{aligned} \tag{18}$$

Inserting Eq. (18) into Eq. (17) and then into Eq. (14), one can finally obtain the expansion of the scattered field $\mathbf{\Psi}^{\mathrm{sca}}$ upon the basis of the QNM field $\tilde{\mathbf{\Psi}}_r$ of the coupled system,

$$\mathbf{\Psi}^{\mathrm{sca}}(\mathbf{r},\omega)=\sum_r\frac{1}{\omega-\tilde{\omega}_r}\frac{\mathbf{h}_r^{\mathrm{T}}\mathbf{b}(\omega)}{\partial\det[\mathbf{A}(\omega)]/\partial\omega|_{\omega=\tilde{\omega}_r}}\tilde{\mathbf{\Psi}}_r(\mathbf{r}), \tag{19}$$

where $\tilde{\mathbf{\Psi}}_r(\mathbf{r})=\mathbf{\Phi}(\mathbf{r})^{\mathrm{T}}\tilde{\mathbf{a}}_r$ is used and the $\mathbf{A}(\infty)^{-1}$ in Eq. (17) is neglected [$\mathbf{A}(\infty)^{-1}=\mathbf{0}$ holds rigorously if $\mathbf{A}(\omega)^{-1}$ is a rational proper-fraction function of $\omega$].

Here we emphasize that the only assumption adopted in the above QCT is that in Eq. (5), the QNMs $\mathbf{\Psi}_{p,m}$ of the $p$th resonator form a complete set of basis in expanding the scattered field $\mathbf{\Psi}_p^{\mathrm{sca}}$ of the $p$th resonator. However, as indicated in Sec. I (paragraph 2), for resonators of analytically-solvable simple geometries [10,19] (such as the present numerical example of 1D slabs), the completeness of physical QNMs (i.e., not-regularized QNMs) holds inside but not outside the resonator. For resonators of general geometries, the QNMs should be regularized (called regularized QNMs, abbreviated as RQNMs) to ensure the completeness in expanding the scattered field both inside and outside the resonator [14,15,21,22,25,26]. In the following subsections B and C, we will provide some theoretical details on the completeness of RQNMs and their incorporation into the QCT.

## B. Completeness of ESC-RQNMs and their incorporation into the QCT

Firstly, we will introduce the definition of ESC-RQNMs along with a theoretical demonstration of their completeness in expanding the scattered field both inside and outside a resonator, which is for the general case that the resonator can be located in an arbitrary inhomogeneous background medium.

According to the field equivalence principle [5,40], the scattered field $\mathbf{\Psi}_p^{\mathrm{sca}}$ outside the $p$th resonator can be regarded as the field radiated in the background medium by an equivalent surface current (ESC) composed of a surface electric current of density $\mathbf{J}_{\mathrm{e},p}(\mathbf{r},\omega)=\hat{\mathbf{n}}\times\mathbf{H}_p^{\mathrm{sca}}(\mathbf{r},\omega)\delta_{S_p}$ and a surface magnetic current of density $\mathbf{J}_{\mathrm{m},p}(\mathbf{r},\omega)=-\hat{\mathbf{n}}\times\mathbf{E}_p^{\mathrm{sca}}(\mathbf{r},\omega)\delta_{S_p}$ on a closed surface $S_p$ encircling the $p$th resonator, where $\hat{\mathbf{n}}$ is the out-pointing unit normal vector on $S_p$, and $\delta_{S_p}$ represents a surface Dirac function on $S_p$ [41]. Then the $\mathbf{\Psi}_p^{\mathrm{sca}}$ outside $S_p$ can be expressed as,

$$\begin{aligned}\mathbf{\Psi}_p^{\mathrm{sca}}(\mathbf{r},\omega)&=\iiint_{R^3}\mathbf{G}^{\mathrm{b}}(\mathbf{r},\mathbf{r}',\omega)\cdot\begin{bmatrix}\hat{\mathbf{n}}\times\mathbf{H}_p^{\mathrm{sca}}(\mathbf{r}',\omega)\delta_{S_p}\\ \hat{\mathbf{n}}\times\mathbf{E}_p^{\mathrm{sca}}(\mathbf{r}',\omega)\delta_{S_p}\end{bmatrix}d^3\mathbf{r}'\\&=\iint_{S_p}\mathbf{G}^{\mathrm{b}}(\mathbf{r},\mathbf{r}',\omega)\cdot\begin{bmatrix}\hat{\mathbf{n}}\times\mathbf{H}_p^{\mathrm{sca}}(\mathbf{r}',\omega)\\ \hat{\mathbf{n}}\times\mathbf{E}_p^{\mathrm{sca}}(\mathbf{r}',\omega)\end{bmatrix}ds',\ \mathbf{r}\in O(S_p),\end{aligned}\tag{20}$$

where $O(S_p)$ denotes the region outside $S_p$, and

$$\mathbf{G}^{\mathrm{b}}(\mathbf{r},\mathbf{r}',\omega)=\begin{bmatrix}\mathbf{G}_{\mathrm{EE}}^{\mathrm{b}}(\mathbf{r},\mathbf{r}',\omega) & \mathbf{G}_{\mathrm{EH}}^{\mathrm{b}}(\mathbf{r},\mathbf{r}',\omega)\\ \mathbf{G}_{\mathrm{HE}}^{\mathrm{b}}(\mathbf{r},\mathbf{r}',\omega) & \mathbf{G}_{\mathrm{HH}}^{\mathrm{b}}(\mathbf{r},\mathbf{r}',\omega)\end{bmatrix}\tag{21}$$

denotes the Green's tensor of the background medium that can be generally inhomogeneous, with $\mathbf{G}_{\mathrm{EE}}^{\mathrm{b}}$, $\mathbf{G}_{\mathrm{EH}}^{\mathrm{b}}$, $\mathbf{G}_{\mathrm{HE}}^{\mathrm{b}}$, and $\mathbf{G}_{\mathrm{HH}}^{\mathrm{b}}$ denoting the electric-field by electric-source, electric-field by magnetic-source, magnetic-field by electric-source, and magnetic-field by magnetic-source Green's tensors, respectively. Under the following assumption (called *Assumption 1*),

The QNMs $\mathbf{\Psi}_{p,m}(\mathbf{r})=[\mathbf{E}_{p,m},\ \mathbf{H}_{p,m}]^{\mathrm{T}}$ of the $p$th resonator form a complete basis in expanding its scattered field $\mathbf{\Psi}_p^{\mathrm{sca}}$ inside and on $S_p$, where the $S_p$ is properly chosen to encircle the $p$th resonator in its near-field region,

One can obtain that Eq. (5) holds inside and on $S_p$, i.e.,

$$\mathbf{\Psi}_p^{\mathrm{sca}}(\mathbf{r},\omega)=\sum_{m=1}^{M_p}\alpha_{p,m}(\omega)\mathbf{\Psi}_{p,m}(\mathbf{r}),\ \mathbf{r}\in I(S_p),\tag{22}$$

where $I(S_p)$ denotes the region inside and on $S_p$, and the $\alpha_{p,m}$ is still given by Eq. (6). The validity of the *Assumption 1* is supported by the numerical and theoretical results reported in the literature [42]. Inserting Eq. (22) into Eq. (20), one can obtain,

$$\begin{aligned}&\mathbf{\Psi}_p^{\mathrm{sca}}(\mathbf{r},\omega)\\&=\iint_{S_p}\mathbf{G}^{\mathrm{b}}(\mathbf{r},\mathbf{r}',\omega)\cdot\begin{bmatrix}\hat{\mathbf{n}}\times\sum_{m=1}^{M_p}\alpha_{p,m}(\omega)\mathbf{H}_{p,m}(\mathbf{r}')\\ \hat{\mathbf{n}}\times\sum_{m=1}^{M_p}\alpha_{p,m}(\omega)\mathbf{E}_{p,m}(\mathbf{r}')\end{bmatrix}ds'\\&=\sum_{m=1}^{M_p}\alpha_{p,m}(\omega)\iint_{S_p}\mathbf{G}^{\mathrm{b}}(\mathbf{r},\mathbf{r}',\omega)\cdot\begin{bmatrix}\hat{\mathbf{n}}\times\mathbf{H}_{p,m}(\mathbf{r}')\\ \hat{\mathbf{n}}\times\mathbf{E}_{p,m}(\mathbf{r}')\end{bmatrix}ds',\ \mathbf{r}\in O(S_p).\end{aligned}\tag{23}$$

Then the $m$th ESC-regularized QNM (ESC-RQNM) of the $p$th resonator is defined as

$$\mathbf{\Psi}_{p,m}^{\mathrm{ESC}}(\mathbf{r},\omega)=\begin{cases}\mathbf{\Psi}_{p,m}(\mathbf{r}),\ \mathbf{r}\in I(S_p),\\ \iint_{S_p}\mathbf{G}^{\mathrm{b}}(\mathbf{r},\mathbf{r}',\omega)\cdot\begin{bmatrix}\hat{\mathbf{n}}\times\mathbf{H}_{p,m}(\mathbf{r}')\\ \hat{\mathbf{n}}\times\mathbf{E}_{p,m}(\mathbf{r}')\end{bmatrix}ds',\ \mathbf{r}\in O(S_p).\end{cases}\tag{24}$$

With this definition, Eq. (22) can be rewritten as,

$$\mathbf{\Psi}_p^{\mathrm{sca}}(\mathbf{r},\omega)=\sum_{m=1}^{M_p}\alpha_{p,m}(\omega)\mathbf{\Psi}_{p,m}^{\mathrm{ESC}}(\mathbf{r},\omega),\ \mathbf{r}\in I(S_p),\tag{25}$$

and Eq. (23) can be rewritten as,

$$\mathbf{\Psi}_p^{\mathrm{sca}}(\mathbf{r},\omega)=\sum_{m=1}^{M_p}\alpha_{p,m}(\omega)\mathbf{\Psi}_{p,m}^{\mathrm{ESC}}(\mathbf{r},\omega),\ \mathbf{r}\in O(S_p).\tag{26}$$

Equations (25) and (26) can be collectively rewritten as,

$$\mathbf{\Psi}_p^{\mathrm{sca}}(\mathbf{r},\omega)=\sum_{m=1}^{M_p}\alpha_{p,m}(\omega)\mathbf{\Psi}_{p,m}^{\mathrm{ESC}}(\mathbf{r},\omega),\ \mathbf{r}\in R^3,\tag{27}$$

where $\alpha_{p,m}$ is still given by Eq. (6). Equation (27) shows that under the *Assumption 1*, the ESC-RQNMs defined by Eq. (24) form a complete basis in expanding the scattered field $\mathbf{\Psi}_p^{\mathrm{sca}}$ both inside and outside the $S_p$ (i.e., throughout the whole space $R^3$). Here note that different from the QNM that corresponds to a complex eigenfrequency and is independent of the real excitation frequency $\omega$, the ESC-RQNM is dependent on $\omega$ via the background Green's tensor $\mathbf{G}^{\mathrm{b}}$.

Secondly, we consider the incorporation of ESC-RQNMs into the QCT, which is performed in a rigorous manner based on the completeness of ESC-RQNMs [Eq. (27)] and the QCT [Eq. (6)] derived from the first principle of Maxwell's equations. Similarly to the derivation of QNM-coupling equations (7), by substituting Eq. (27) into Eq. (6) to replace the $\mathbf{E}_q^{\mathrm{sca}}$ by $\sum_{n=1}^{M_q}\alpha_{q,n}(\omega)\mathbf{E}_{q,n}^{\mathrm{ESC}}(\mathbf{r},\omega)$, one can obtain the ESC-RQNM coupling equations, which have the same form of the QNM-coupling equations (7), but with revised definition,

$$\begin{aligned}&\kappa_{(p,m),(q,n)}(\omega)\\&=-\frac{\omega}{F_{p,m}}\iiint_{V_p}\mathbf{E}_{p,m}(\mathbf{r})\cdot\Delta\boldsymbol{\varepsilon}_p(\mathbf{r},\omega)\cdot\mathbf{E}^{\mathrm{ESC}}_{q,n}(\mathbf{r},\omega)d^3\mathbf{r},\end{aligned}\tag{28}$$

where the $\mathbf{E}_{p,m}$ can also be replaced by $\mathbf{E}^{\mathrm{ESC}}_{p,m}$ according to Eq. (24) and $V_p \subseteq I(S_p)$, while the pseudoenergy $F_{p,m}$ has no change. In summary, compared with the QCT using QNMs, the only change in the QCT using ESC-RQNMs is that Eqs. (5) and (8) are changed to be Eqs. (27) and (28), respectively [43].

One difficulty caused by the replacement of QNMs with ESC-RQNMs is that, for solving Eq. (13) to obtain the QNMs of the coupled system, the $\kappa_{(p,m)(q,n)}$ is dependent on the background Green's tensor $\mathbf{G}^{\mathrm{b}}$ which is typically an exponentially nonlinear function of frequency $\omega$, so that Eq. (13) becomes a transcendental equation with an exponential complexity. Consequently, it is difficult to obtain all the solutions of Eq. (13), and commonly only a part of its solutions close to the iterative initial values can be obtained by using an iterative method [44]. If using the PML-RQNMs (see the next subsection C), as explained in Ref. [6], Eq. (12) can reduce to a polynomial matrix eigenvalue problem, for which all the solutions can be obtained with well-developed algorithms [38]. However, for solving the scattered field excited by a background field, which relies on solving the set of linear equations (10), no extra difficulty is caused by the replacement of QNMs with ESC-RQNMs.

The above incorporation of the ESC-RQNMs into the QCT has been partially discussed in previous literatures [5,25,26]. In Ref. [5] [see Eqs. (119) and (120) therein], some QNMs of the coupled system are solved by the QCT incorporating a regularized field outside the resonator, which is methodologically equivalent to the QCT [Eqs. (12) and (13)] using ESC-RQNMs (although the concept of ESC-RQNM is not explicitly proposed in Ref. [5]), and the considered numerical example is a coupled system composed of two identical separated 1D slabs (see Fig. 25 therein). In Refs. [25,26] by some of the authors of this paper, the scattered field of the coupled system excited by a background field is solved [i.e., solving Eq. (10)] by using the ESC-RQNMs, which is demonstrated by an efficient modeling of the numerically difficult example of a large-scale array of optical nanoresonators, where the definition of ESC-RQNMs and a theoretical demonstration of their completeness are provided for resonators in a homogeneous background medium [25] (or on a flat-interface substrate [26]) via a plane-wave expansion (or expansion in terms of radially-propagating waveguide modes) which is logically equivalent to the background Green's tensor in Eq. (20).

## C. Completeness of PML-RQNMs and their incorporation into the QCT

Firstly, we will introduce the definition of PML-RQNMs along with some remarks on their completeness in expanding the scattered field both inside and outside a resonator. For solving the QNMs numerically as the eigensolutions of source-free Maxwell's equations, the PMLs are introduced as a complex coordinate transformation (or equivalently, as an effective absorbing medium) to handle the outgoing-wave condition at infinity satisfied by the QNMs, so as to map the infinite physical space to a finite computational domain and a surrounding PML domain [14,45] (see SM Sec. S3 B [31] for more details). This use of PMLs for solving QNMs brings two benefits. The first benefit is that the outgoing far field of QNMs is transformed from exponential divergence in physical space to exponential decay in the PMLs, which then enables a proper normalization of the QNMs by using the exponentially decaying field in the PMLs [12]. Therefore, the QNMs solved numerically via the introduction of PMLs are called PML-regularized QNMs (PML-RQNMs) [15]. The second benefit is that due to the discretization of Maxwell's equations over the finite computational domain and the surrounding PML domain, the resultant finite degrees of freedom (i.e., number of unknowns) lead to a finite number of the solved PML-RQNMs, which then enables a numerical completeness of the solved PML-RQNMs within the computational domain (containing both the inside and the outside of a resonator) [14,21,22].

Due to the introduction of PMLs and discretization of Maxwell's equations, the solved PML-RQNMs can be classified into physical PML-RQNMs and numerical PML-RQNMs [14,15,35]. The physical PML-RQNMs have nearly the same eigenfrequencies and electromagnetic fields (out of the PMLs) as those of the physical QNMs, and are almost independent of the parameters of PMLs. Here the physical QNMs (or called not-regularized QNMs) are defined as the eigensolutions of the exact Maxwell's equations without discretization. The numerical PML-RQNMs exhibit significant deviation from the physical QNMs, and can be further classified into transitional PML-RQNMs and purely numerical PML-RQNMs. The transitional PML-RQNMs originate from the physical QNMs that are significantly perturbed by the imperfectness of PMLs (electromagnetic field not well attenuated) and discretization of Maxwell's equations (electromagnetic field not well resolved), and will approach the physical QNMs when improving the PMLs and discretization [14,15,35]. The purely numerical PML-RQNMs have no obvious correlation with the physical QNMs, and strongly depend on the PML parameters or the discretization of Maxwell's equations. Note that the above classification of PML-RQNMs only makes sense approximately, since no sharp boundary exists between the different types of PML-RQNMs.

Secondly, we consider the incorporation of PML-RQNMs into the QCT. Based on the completeness of PML-RQNMs, Eq. (5) holds within the computational domain truncated by the PMLs after replacing the QNMs $\mathbf{\Psi}_{p,m}$ of the $p$th resonator by the corresponding PML-RQNMs $\mathbf{\Psi}_{p,m}^{\text{PML}}$. After this replacement, the QNM-coupling equations (7) remain valid as long as the computational domain of the $q$th resonator is large enough to contain the $p$th resonator (for any $p$, $q$=1, 2, …, $P$), which is required as well for calculating the QNM coupling coefficient $\kappa_{(p,m),(q,n)}$ defined in Eq. (8). Therefore, the incorporation of PML-RQNMs leads to no change of the QCT except for replacing the QNMs by PML-RQNMs. Specifically, for the $\kappa_{(p,m),(q,n)}$ defined in Eq. (8), the incorporated PML-RQNMs are independent of the excitation frequency $\omega$, which is like the case of QNMs but different from the case of ESC-RQNMs [Eq. (28)]. This brings the benefit that for the QCT using PML-RQNMs, Eq. (12) can still reduce to a polynomial matrix eigenvalue problem, for which all the eigen solutions can be solved [6] with well-developed algorithms [38]. However, this benefit is unattainable for the QCT using ESC-RQNMs, as explained in Sec. II B following Eq. (28).

On the other hand, it should be noted that as the distance between two resonators is much larger than the wavelength, such as the situation encountered in a large-scale array of resonators [25,26], the calculation of coupling coefficient $\kappa_{(p,m),(q,n)}$ between the PML-RQNMs of the two resonators requires a large PML-truncated computational domain that should contain both of the two resonators, which will result in a large computational load for solving the PML-RQNMs. This weak point, however, can be avoided for the QCT using ESC-RQNMs when the background Green's tensor has an analytical expression, for instance, for the background medium of homogeneous free space [25] or flat-interface substrate [26].

# III. NUMERICAL RESULTS

To implement a strict numerical test for the impact of mode completeness on the accuracy of QCT, it is required to solve a large amount of RQNMs of each resonator so as to form a virtually complete basis, which is difficult for general 3D structures. Here we consider a numerical example of coupled resonators of 1D slabs, for which all the RQNMs can be solved either analytically (for ESC-RQNMs) or numerically (for PML-RQNMs). In order to provide a strong test of the impact of mode completeness on the accuracy of the QCT, we consider the extreme coupling case of direct contact of two slabs [as shown in Fig. 1(a)], for which the not-regularized QNMs (i.e., physical QNMs) cannot ensure the completeness (to be shown in Sec. III A), and a large number of ESC-RQNMs or PML-RQNMs are required to ensure the completeness (Sec. III B).

## A. Results of the QCT using physical QNMs (i.e., not-regularized QNMs)

As shown in Fig. 1(a), the coupled system is composed of two identical 1D slabs (with a width $d$) of direct contact in a homogeneous background medium. Thus the coupled system is also a slab (width 2$d$). Here we assume that the coupled system is made of nonmagnetic and isotropic media without frequency dispersion, and that the refractive indices of the background medium and the slabs are $n_1$=1 and $n_2$=1.5, respectively. The QNMs of this simple structure can be solved analytically and are called physical QNMs or not-regularized QNMs. For the single slab of width $d$, the eigenfrequencies $\omega=\omega_{\text{exact}}$ of the physical QNMs are,

$$\omega_{\text{exact}} = \omega_{+,m} = \frac{c}{d}\frac{1}{n_2}(2m\pi + i\ln|r_{21}|), \tag{29a}$$

$$\omega_{\text{exact}} = \omega_{-,m} = \frac{c}{d}\frac{1}{n_2}[(2m-1)\pi + i\ln|r_{21}|], \tag{29b}$$

where $m$ takes integers, $\omega_{+,m}$ and $\omega_{-,m}$ represent the eigenfrequencies of symmetric and anti-symmetric QNMs, respectively, and $c$ is the light speed in the vacuum. The $r_{ij}=(n_i-n_j)/(n_i+n_j)$ is the reflection coefficient of a normally incident plane wave propagating in $z$-direction from medium $i$ to medium $j$. The electric field $\mathbf{E}_{\pm,m}(z)=E_{\pm,m,x}(z)\,\hat{\mathbf{x}}$ of the $m$th-order symmetric (+) or anti-symmetric (–) physical QNM can be expressed as,

$$E_{\pm,m,x}(z) = \begin{cases} b_{1,\pm,m}\exp[-ik_{\pm,m}n_1(z-z_1)],\ z<z_1, \\ a_{2,\pm,m}\exp[ik_{\pm,m}n_2(z-z_1)] \\ +b_{2,\pm,m}\exp[-ik_{\pm,m}n_2(z-z_2)],\ z_1<z<z_2, \\ a_{1,\pm,m}\exp[ik_{\pm,m}n_1(z-z_2)],\ z>z_2, \end{cases} \tag{30}$$

where $z_1$=0 and $z_2$=$d$ are the $z$ coordinates of the two boundaries of the single slab, $a_{1,\pm,m}=\pm t_{21}/r_{21}$, $b_{1,\pm,m}=t_{21}/r_{21}$, $a_{2,\pm,m}$=1, $b_{2,\pm,m}$=±1, $k_{\pm,m}=\omega_{\pm,m}/c$, and $t_{ij}=2n_i/(n_i+n_j)$ represents the transmission coefficient of a normally incident plane wave propagating in $z$-direction from medium $i$ to medium $j$. The magnetic field $\mathbf{H}_{\pm,m}(z)=H_{\pm,m,y}(z)\,\hat{\mathbf{y}}$ can be obtained through $H_{\pm,m,y}(z)=(i\omega\mu_0)^{-1}dE_{\pm,m,x}(z)/dz$, with $\mu_0$ being the permeability in the vacuum. For the two coupled slabs, the QNM eigenfrequencies $\omega=\tilde{\omega}_{\text{exact}}=\tilde{\omega}_{\pm,m}$ are simply Eq. (29) with $d$ replaced by 2$d$, and the QNM electric fields are simply Eq. (30) with $z_1=-d$ and $z_2=d$ denoting the boundary coordinates of the coupled slabs.

As shown in Fig. 1(b), the eigenfrequencies of the two coupled slabs obtained from Eq. (29) (with $d$ replaced by 2$d$) are compared with the results obtained with the QCT [Eq. (13)]. In the QCT, $M_p=M$=202, 402, 802 physical QNMs with the minimum $|\text{Re}(\omega_{p,m})|$ are retained for each slab. Figure 1(b) shows that the results of QCT almost do not change if further increasing $M$ from 402 to 802, but such unchanged results are far from the exact solution [17]. The results of QCT in Fig. 1(b) are identical to the results shown by the circles in Fig. 1 of Ref. [17], except that the latter are obtained using 3200 QNMs.

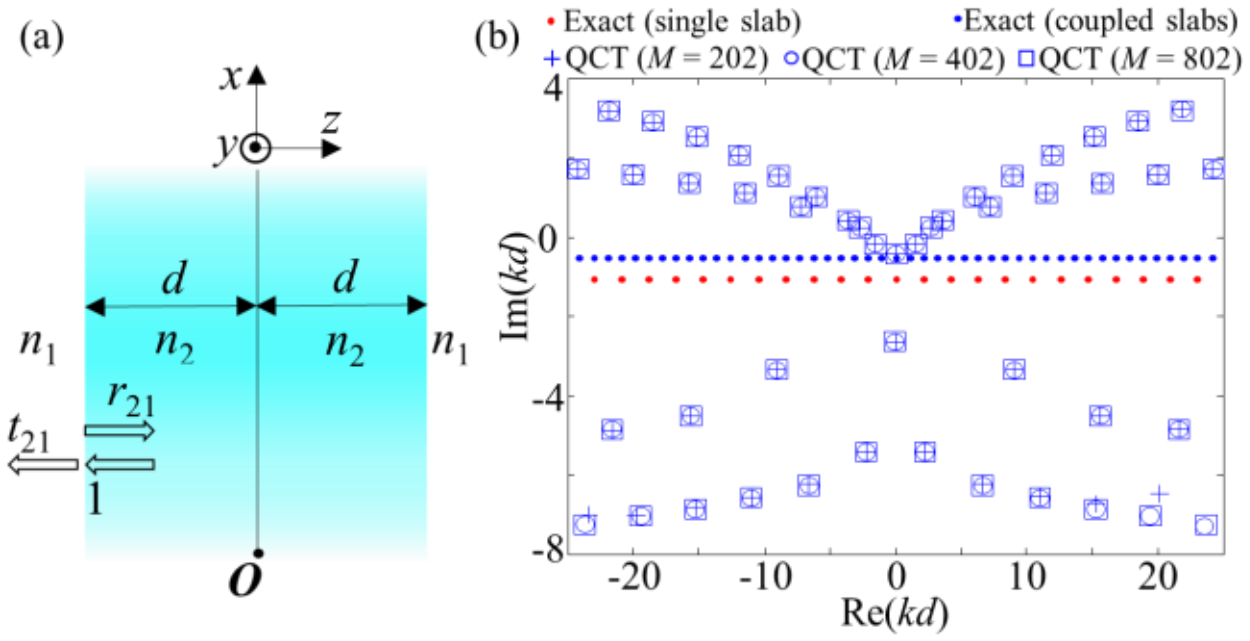

Fig. 1 (a) Schematic of the two coupled identical slabs. The origin $O$ of coordinates is set on the interface between the two slabs. (b) Results of the eigenfrequencies $\omega$. Here $k=\omega/c$ with $c$ being the speed of light in vacuum. The red and blue marks represent the eigenfrequencies of single slab and coupled slabs, respectively. The blue dots represent the analytical exact solution of Eq. (29) (with $d$ replaced by $2d$). The blue pluses, circles and squares represent the results obtained with the QCT using $M_p=M=202$, 402 and 802 physical QNMs of a single slab, respectively.

In addition to the above source-free eigenproblem, we also consider a source-excitation scattering problem in the numerical test. The two coupled slabs (with boundary coordinates $z_1=-d$ and $z_2=d$) are excited by an $x$-polarized plane wave incident from the left side of the slabs with the electric field $\mathbf{E}^{\mathrm{b}}(z)=\exp[ikn_1(z-z_1)]\,\hat{\mathbf{x}}$ . For the scattered field defined as the total field minus the incident field (with and without the presence of the coupled slabs, respectively), the exact solution of its electric field $\mathbf{E}^{\mathrm{sca}}= E^{\mathrm{sca}}_{x,\mathrm{exact}}\hat{\mathbf{x}}$ is,

$$E^{\mathrm{sca}}_{x,\mathrm{exact}}(z)=\begin{cases} R\exp[-ikn_1(z-z_1)],\ z<z_1, \\ a\exp[ikn_2(z-z_1)]-\exp[ikn_1(z-z_1)] \\ +b\exp[-ikn_2(z-z_2)],\ z_1<z<z_2, \\ T\exp[ikn_1(z-z_2)]-\exp[ikn_1(z-z_1)],\ z>z_2, \end{cases} \tag{31}$$

where $k=\omega/c$ and

$$a=\frac{t_{12}}{1-u^2r_{21}^2},\ b=\frac{t_{12}ur_{21}}{1-u^2r_{21}^2},\ T=\frac{ut_{12}t_{21}}{1-u^2r_{21}^2},\ R=\frac{t_{12}(1+u^2r_{21})}{1-u^2r_{21}^2}-1,$$

with $u=\exp[ikn_2(z_2-z_1)]$. Then the scattered magnetic field can be obtained as $\mathbf{H}^{\mathrm{sca}}= H^{\mathrm{sca}}_{y,\mathrm{exact}}\hat{\mathbf{y}}$ with

$$H^{\mathrm{sca}}_{y,\mathrm{exact}}=(i\omega\mu_0)^{-1}dE^{\mathrm{sca}}_{x,\mathrm{exact}}(z)/dz.$$

The results of scattered field at excitation frequency $\omega=\mathrm{Re}(\tilde{\omega}_{+,1})$ are shown in Fig. 2. The red-dotted and blue-dashed curves are obtained from the exact solution of Eq. (31) and from Eq. (11) of QCT, respectively. For the QCT, $M_p=M=802$ physical QNMs with the minimum $|\mathrm{Re}(\omega_{p,m})|$ are retained for each slab, and we have confirmed that further increasing $M$ hardly changes the results. Again, the predictions of QCT are far from the exact solution.

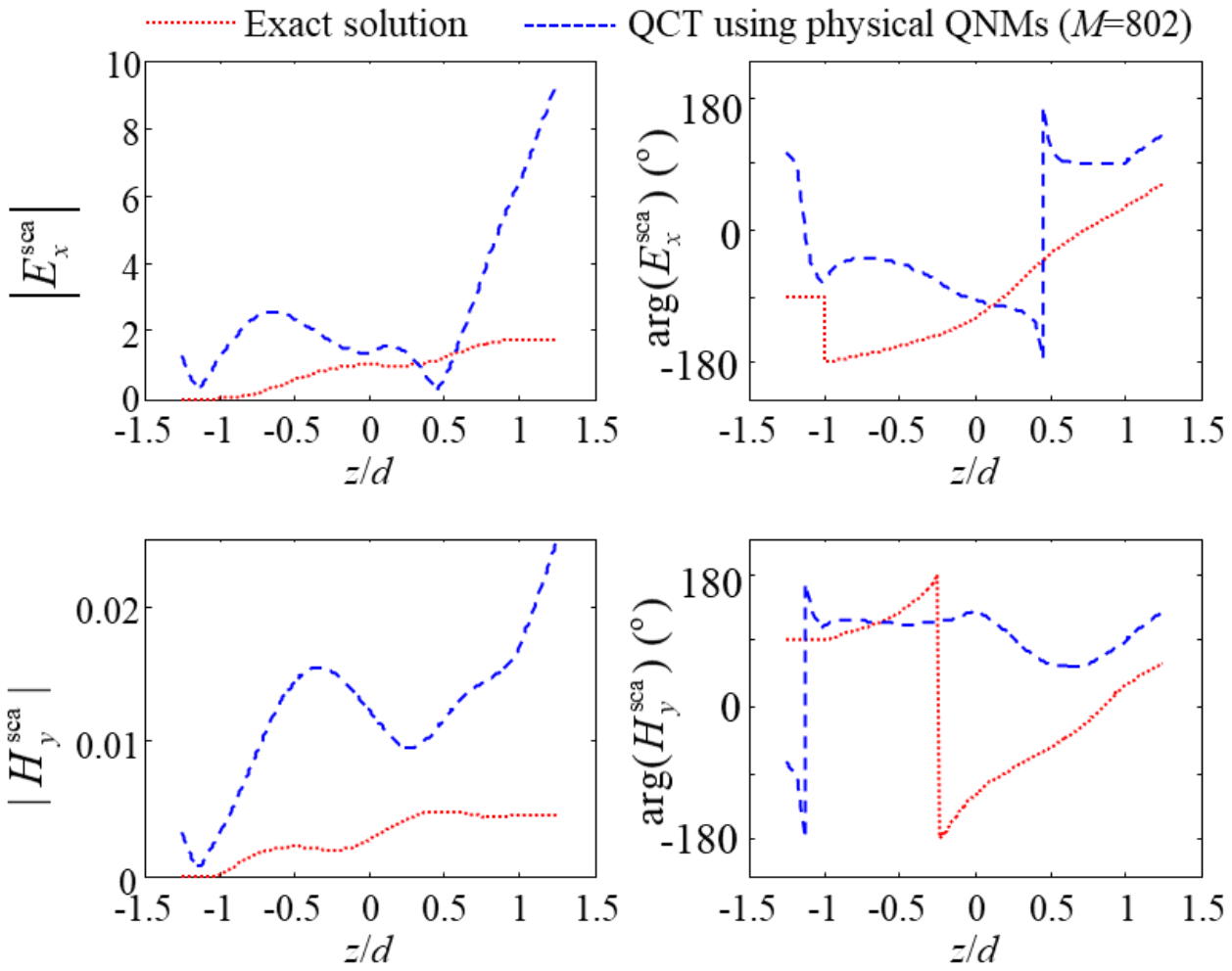

Fig. 2 Results of the scattered field of the coupled slabs illuminated by a plane wave with excitation frequency $\omega=\mathrm{Re}(\tilde{\omega}_{+,1})$. The red-dotted and blue-dashed curves show the results obtained from analytical exact solution and QCT, respectively.

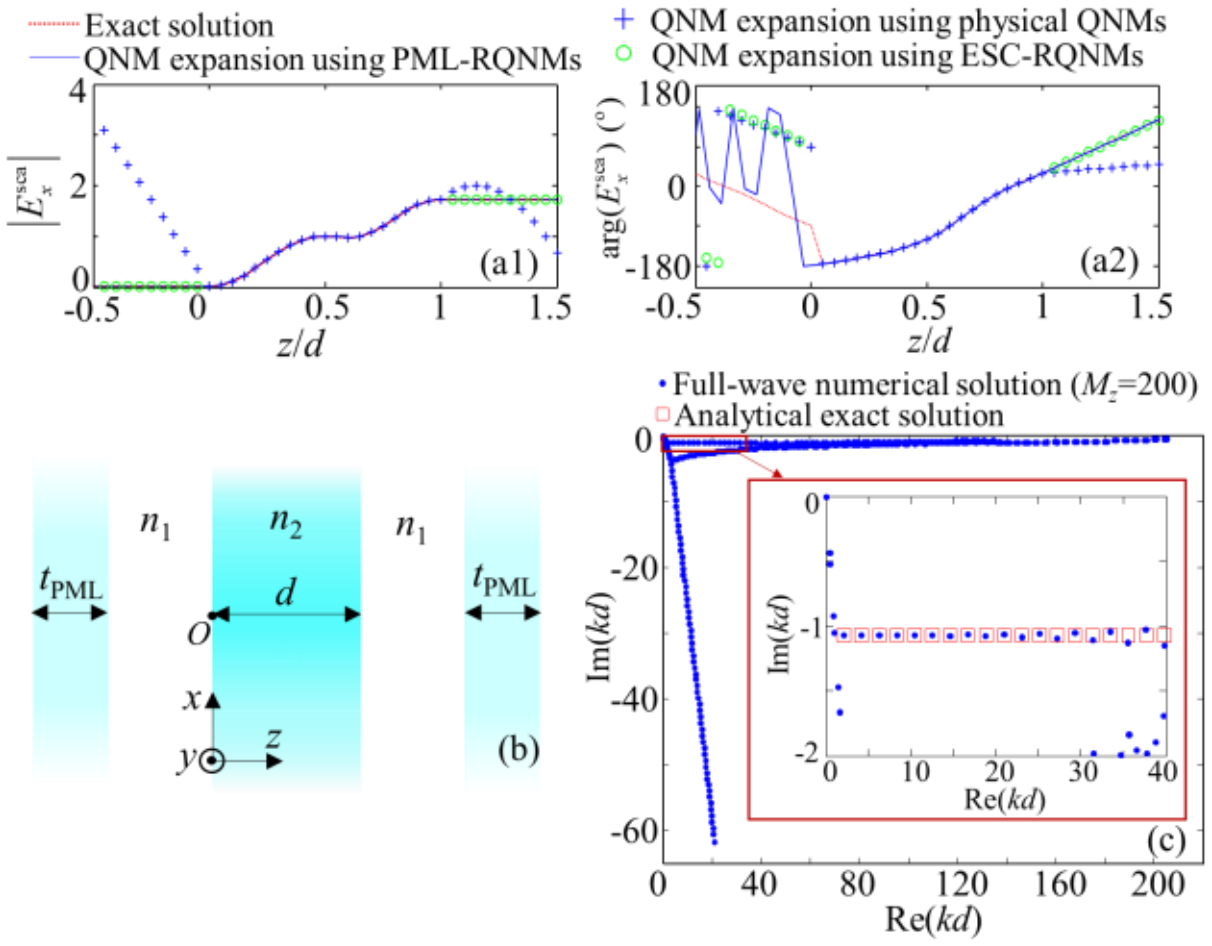

Fig. 3 (a) Results of the scattered field of a single slab (boundaries at $z=0$ and $d$) excited by a plane wave at frequency $\omega=\mathrm{Re}(\omega_{1,+})$. (b) Schematic of the single slab surrounded by two PMLs (thickness $t_{\mathrm{PML}}$). (c) QNM eigenfrequencies of the single slab. The red squares and blue dots are the results of the analytical exact solution and the full-wave numerical solution of PWEM (with a truncated harmonic order $M_z=200$), respectively.

The above numerical tests in Figs. 1 and 2 consistently show that the QCT only using physical QNMs exhibits a large error for this numerical example of coupled slabs, which is expected to result from the incompleteness of the physical QNMs (i.e., not-regularized QNMs), as also indicated in Ref. [17]. To confirm this expectation, we calculate the scattered field of a single slab using the QNM expansion formalism [i.e., Eqs. (5) and (6) with the scatterer number $P$=1] where only the physical QNMs are retained [$M_p$=$M$=802 modes with the minimum $|\mathrm{Re}(\omega_{p,m})|$]. As shown in Fig. 3(a), compared with the exact solution (red dotted curves), the result of QNM expansion (blue pluses) is accurate inside the slab but exhibits a large error outside the slab. This confirms the expectation that the not-regularized QNMs are complete inside the slab but not outside the slab [10], which results in the large error of QCT. Here the exact solution of the scattered field of a single slab is simply Eq. (31) with $z_1$=0 and $z_2$=$d$ denoting the boundary coordinates of the single slab.

## B. Results of the QCT using ESC-RQNMs or PML-RQNMs as a complete basis

In this subsection, we will first provide a numerical test of the completeness of ESC-RQNMs and PML-RQNMs in expanding the source-excited scattered field of a single 1D slab (Sec. III B1), and then provide the results of the QCT using ESC-RQNMs or PML-RQNMs as a complete basis in predicting the source-free QNMs (Sec. III B2) and source-excited scattered field (Sec. III B3) of two coupled slabs.

### 1. Numerical test of the completeness of ESC-RQNMs and PML-RQNMs

Firstly, we will explain the calculation of the ESC-RQNMs and then provide a numerical test of the completeness of ESC-RQNMs in expanding the source-excited scattered field of a single 1D slab. For the ESC-RQNMs of a single 1D slab, the region $I(S_p)$ in Eq. (24) is chosen as the interior of the slab, i.e., $z_1 \leqslant z \leq z_2$ with $z_1$=0 and $z_2$=$d$ [see the coordinate system in Fig. 3(b)]. Then for $z_1 \leq z \leq z_2$, the ESC-RQNMs are defined as the physical QNMs, i.e., $\mathbf{\Psi}^{\mathrm{ESC}}_{\pm,m}(z,\omega)=\mathbf{\Psi}_{\pm,m}(z)$ as given by Eq. (30). Out of $z_1 \leq z \leq z_2$, Eq. (24) shows that the electric fields $\mathbf{E}^{\mathrm{ESC}}_{\pm,m}(z,\omega)=\hat{\mathbf{x}}E^{\mathrm{ESC}}_{\pm,m,x}(z,\omega)$ of the ESC-RQNMs are given by,

$$E^{\mathrm{ESC}}_{\pm,m,x}(z,\omega)=\begin{cases} b_{1,\pm,m}\exp[-ikn_1(z-z_1)],\ z<z_1, \\ a_{1,\pm,m}\exp[ikn_1(z-z_2)],\ z>z_2, \end{cases} \tag{32}$$

where $k$=$\omega/c$ with $\omega$ being the real-valued excitation frequency, and the analytical expressions for the Green's tensor of homogeneous background medium and for the physical QNMs are used. Equation (32) appears to be the same as Eq. (30) for the physical QNM except that $k_{\pm,m}$ is replaced by $k$. The magnetic field $\mathbf{H}^{\mathrm{ESC}}_{p,m}(z,\omega)=\hat{\mathbf{y}}H^{\mathrm{ESC}}_{p,m,y}(z,\omega)$ of the ESC-RQNMs can be obtained through $H^{\mathrm{ESC}}_{p,m,y}(z)=(i\omega\mu_0)^{-1}\,dE^{\mathrm{ESC}}_{p,m,x}(z)/dz$.

As shown in Fig. 3(a), the scattered field both inside and outside the slab calculated by the QNM expansion using ESC-RQNMs [green circles, using $M_p$=$M$=802 modes with the minimum $|\mathrm{Re}(\omega_{p,m})|$] agree well with the exact solution [red dotted curves, given by Eq. (31) with $z_1$=0 and $z_2$=$d$], which then verifies the completeness of ESC-RQNMs.

Secondly, we will explain the method for calculating the PML-RQNMs and then provide a numerical test of the completeness of PML-RQNMs in expanding the source-excited scattered field of a single 1D slab. For calculating the PML-RQNMs, we propose a full-wave numerical plane-wave expansion method (PWEM), which is simply an application of the method in Ref. [32] to the special case of 1D structures but further incorporates the PMLs. As illustrated in Fig. 3(b), the PML is introduced as a complex coordinate transformation,

$$Z=f(z')=\begin{cases} z_L+(z'-z_L)\,f_{\mathrm{PML}},\ \text{for } z_L-t_{\mathrm{PML}}\le z'\le z_L, \\ z',\ \text{for } z_L\le z'\le z_R, \\ z_R+(z'-z_R)\,f_{\mathrm{PML}},\ \text{for } z_R\le z'\le z_R+t_{\mathrm{PML}}, \end{cases} \tag{33}$$

where $Z$ represents an extension of the real-valued $z$ coordinate to complex plane (denoted as $z \rightarrow Z$), $z'$ is a real-valued numerical coordinate, $z_L$ and $z_R$ are the $z$-coordinates of left and right inner boundaries of the PML, respectively, $t_{\mathrm{PML}}$ is the thickness of the PML, and $f_{\mathrm{PML}}$ takes a complex value with a positive imaginary part as required for attenuating the outgoing wave [15]. For solving the PML-RQNMs of the single slab by using the PWEM, we set $t_{\mathrm{PML}}$=$d$, $f_{\mathrm{PML}}$=1+3$i$, $z_L$=$z_1$−1.5$d$ and $z_R$=$z_2$+1.5$d$, with $z_1$=0 and $z_2$=$d$ denoting the boundary coordinates of the single slab. By using the PWEM, the PML-RQNMs can be obtained by solving a matrix eigenvalue problem, whose eigenvalues and eigenvectors give the eigenfrequencies and electromagnetic fields of PML-RQNMs, respectively. All the eigen solutions of the matrix eigenvalue problem are solved and used in the following numerical test, and are expected to form a virtually complete basis of PML-RQNMs. More details on the implementation of PWEM are provided in SM Sec. S2 [31].

By using the PWEM, the solved eigenfrequencies of a single slab are central symmetric with respect to the origin [as shown in the inset of Fig. S1(a)], which is due to the fact that the medium permittivity and the PMLs are both independent of frequency (see a detailed explanation in SM Sec. S3 D [31]). So only the eigenfrequencies with non-negative real parts are displayed in Fig. 3(c). Besides, this spectral central symmetry of PML-RQNMs is different from the spectral mirror symmetry of the physical QNMs shown by Eq. (29), i.e., a mirror symmetry of eigenfrequencies with respect to the imaginary axis. This difference is due to the fact that only the PML-RQNMs (denoted by PML-RQNM$^{+}$) for expanding the positive-frequency components [denoted by $\mathbf{\Psi}(\omega>0)$] of electromagnetic field are solved by the PWEM. And if further considering the PML-RQNMs (denoted by PML-RQNM$^{-}$) for expanding the negative-frequency components [denoted by $\mathbf{\Psi}(\omega<0)$] of electromagnetic field, which is necessary for constructing the electromagnetic field in time domain, the union of PML-RQNM$^+$ and PML-RQNM$^-$ then satisfies the spectral mirror symmetry of the physical QNMs (further detailed explanations are provided in SM Secs. S3 A-S3 C [31]).

As shown by the local zoom in the inset of Fig. 3(c), a part of the solved PML-RQNMs have almost the same eigenfrequencies as the analytical exact solutions (i.e., physical QNMs), and thereby belong to the physical PML-RQNMs (see the discussion on the classification of PML-RQNMs in Sec. II C). With the increase of Re($k$), the PML-RQNMs exhibit eigenfrequencies gradually deviating from those of physical QNMs, and thus belong to transitional PML-RQNMs. As shown in Fig. S1(a) in SM Sec. S2 [31], these transitional PML-RQNMs will become physical PML-RQNMs by increasing the truncated harmonic order $M_z$ of PWEM. In this paper, we focus on the eigenfrequencies with Re($kd$)<25, for which $M_z$=200 is large enough. Besides, a large number of eigenfrequencies locate at an oblique line, which belong to the purely numerical PML-RQNMs [14,15,21,35].

As shown in Fig. 3(a), the results of QNM expansion using PML-RQNMs (blue solid curves) are in good agreement with the exact solution (red dotted curves) in predicting the scattered field both inside and outside the single slab, which verifies the completeness of the PML-RQNMs. Here all the solved PML-RQNMs (including physical, transitional and purely numerical PML-RQNMs) are used, with a mode number of $2(2M_z+1)$ for a truncated harmonic order $M_z$=200.

In summary, we have numerically verified the completeness of the ESC-RQNMs and the PML-RQNMs in expanding the source-excited scattered field of a single slab. Resultantly, it is expected that the QCT will achieve a high accuracy if using enough number of ESC-RQNMs or PML-RQNMs as a virtually complete basis, which will be tested numerically in the following two subsections.

**2. Results of the QCT in predicting the source-free QNMs of the two coupled slabs**

Figure 4(a) shows the eigenfrequencies of the QNMs of the two coupled slabs. For the QNMs of the coupled slabs located in an infinite background medium [as shown by the lower inset in Fig. 4(a)], the predictions of the QCT using enough number [$M_p$=$M$=802 modes with the minimum $|\mathrm{Re}(\omega_{p,m})|$] of ESC-RQNMs of a single slab as a virtually complete basis (blue squares) are in agreement with the analytical exact solution [red dots, given by Eq. (29) with $d$ replaced by $2d$]. As mentioned in Sec. II B, an exponentially nonlinear eigenvalue problem needs to be solved in the QCT using ESC-RQNMs, and an iterative linear interpolation method [44] is used here to solve the QNMs one by one.

For the PML-RQNMs of the coupled slabs surrounded by PMLs [as shown by the upper inset in Fig. 4(a)], the predictions of QCT using all the solved PML-RQNMs (mode number $M$=802) of a single slab as a virtually complete basis (blue circles) agree well with the full-wave PWEM results (blue pluses, truncated harmonic order $M_z$=400). Here for solving the PML-RQNMs of the coupled slabs by using the PWEM, we set $t_{\mathrm{PML}}$=$d$, $f_{\mathrm{PML}}$=1+3$i$, $z_L$=$z_1$−$d$, and $z_R$=$z_2$+$d$ in Eq. (33), where $z_1$=−$d$ and $z_2$=$d$ denote the boundary coordinates of the coupled slabs.

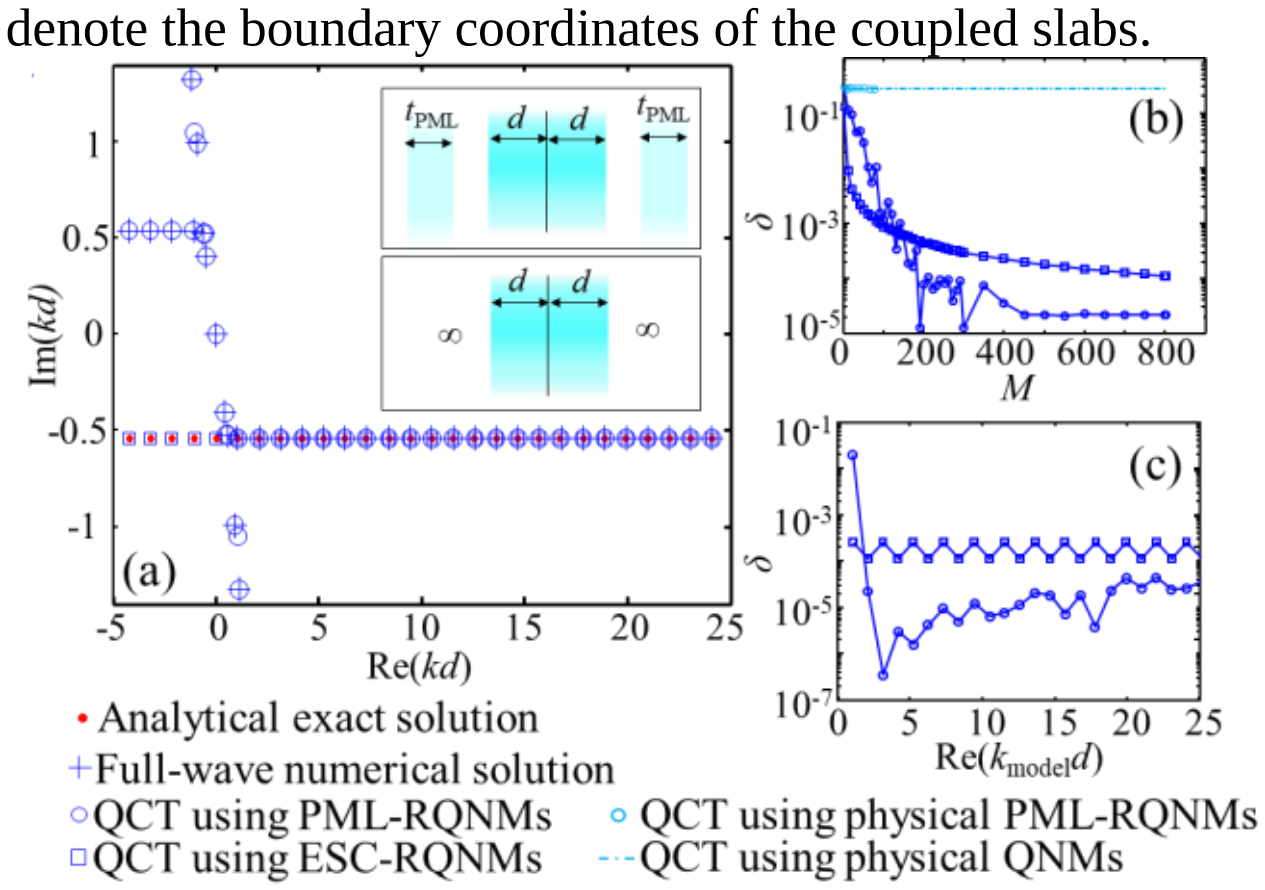

Fig. 4 (a) Eigenfrequencies $\omega$ of the QNMs of the two coupled slabs ($k$=$\omega$/$c$, $c$ being the light speed in the vacuum). For the coupled slabs surrounded by PMLs (upper inset), the blue pluses and blue circles show the results obtained with the full-wave PWEM and the QCT using $M$=802 PML-RQNMs, respectively. For the coupled slabs located in an infinite background medium (lower inset), the red dots and blue squares show the results obtained with the analytical exact solution and the QCT using $M$=802 ESC-RQNMs, respectively. (b) Convergence curves of the QCT in predicting the eigenfrequency $\tilde{\omega}_{+,1}$ of the 1st-order symmetric QNM, where the relative error $\delta$ is plotted as a function of the mode number $M$. (c) Relative error $\delta$ of the QCT in predicting the eigenfrequencies satisfying 0<Re($kd$)<25. The curves in (b) and (c) follow the specifications in (a), except that in (b), the light blue circles and dash-dot curve show the results of the QCT using only physical PML-RQNMs and only physical QNMs (i.e., not-regularized QNMs), respectively.

Figure 4(b) shows the convergence of the eigenfrequency of the 1st-order symmetric QNM [corresponding to the physical QNM with eigenfrequency $\tilde{\omega}_{+,1}$ as defined after Eq. (30) for the coupled slabs], which is obtained by gradually increasing the number $M$ of used modes (sorted with an increasing $|\omega_{p,m}-\omega_{\mathrm{exact}}|$). Here a relative error is defined as $\delta=|\omega_{\mathrm{QCT}}-\omega_{\mathrm{exact}}|/|\omega_{\mathrm{exact}}|$, where $\omega_{\mathrm{QCT}}$ represents the eigenfrequency predicted by the QCT, and $\omega_{\mathrm{exact}}$ represents the exact solution [given by Eq. (29) with $d$ replaced by $2d$]. For the QCT using ESC-RQNMs or PML-RQNMs (including both physical and numerical PML-RQNMs), a good convergence can be observed in Fig. 4(b), for instance, the $\delta$ decreasing down to $1\times10^{-4}$ for $M$=802

ESC-RQNMs, or to $2\times10^{-5}$ for $M$=500 PML-RQNMs. But for the QCT using only physical PML-RQNMs (light blue circles, $M$=50) or only physical QNMs (light blue dash-dot curve, $M$=802), the $\delta$ saturates at a large value of about 0.28 and no convergence is observed.

Figure 4(c) shows the relative error $\delta$ of the QNMs corresponding to some other physical QNMs [with eigenfrequencies satisfying 0<Re($kd$)<25] predicted by the QCT using ESC-RQNMs (blue squares, $M$=802) or PML-RQNMs (blue circles, $M$=802). Again, a high accuracy can be observed. As shown in Fig. S4(a) in SM Sec. S4 [31], by increasing the imaginary part of $f_{\mathrm{PML}}$ [see Eq. (33)] to enhance the decay of PML-RQNM field in the PMLs, the convergence of the QCT using PML-RQNMs can be further improved.

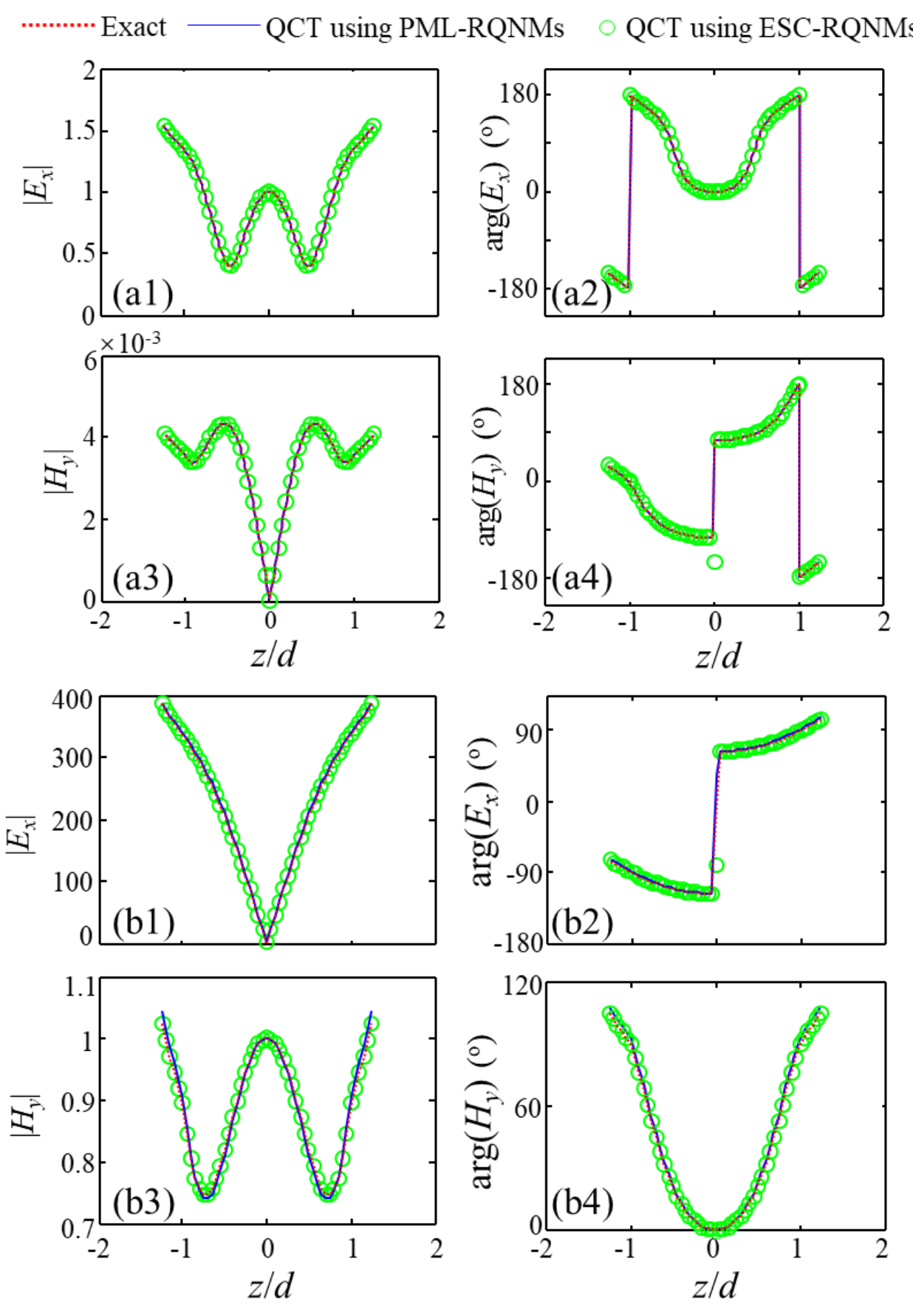

Fig. 5 (a) Electromagnetic field of the 1st-order symmetric QNM of the coupled slabs, which is normalized to have $E_x(z=0)=1$. The red dot lines show the exact solution given by Eq. (30) for coupled slabs. The blue solid lines represent the results of the QCT using all the solved $M$=802 PML-RQNMs of a single slab. The green circles are the results of the QCT using $M$=802 ESC-RQNMs with the smallest $|\mathrm{Re}(\omega_{p,m})|$. (b) The same as (a), but for the 1st-order anti-symmetric QNM which is normalized to have $H_y(z=0)=1$.

The electromagnetic fields of the 1st-order symmetric and anti-symmetric QNMs (corresponding to physical QNMs with eigenfrequencies $\tilde{\omega}_{+,1}$ and $\tilde{\omega}_{-,1}$, respectively) are shown in Fig. 5, where $z$=0 is the center of the two coupled slabs. The results show that the predictions by the QCT using ESC-RQNMs [green circles, using $M$=802 modes with the minimum $|\mathrm{Re}(\omega_{p,m})|$] or all the solved PML-RQNMs (blue solid curves, mode number $M$=802) are both consistent with the exact solutions [red dot curves, given by Eq. (30) for coupled slabs].

For the QCT using PML-RQNMs, Fig. 4(a) shows that the physical QNMs (red dots) with a negative real part of eigenfrequency cannot be predicted. This is because the QCT only uses PML-RQNMs$^{+}$ [see the relevant definitions after Eq. (33)] of a single slab, so that the QCT can only predict the PML-RQNMs$^{+}$ of the coupled slabs, which do not include the physical QNMs with a negative real part of eigenfrequency. As shown in Fig. S5 in SM Sec. S5 [31], if the QCT uses PML-RQNMs$^{-}$ of a single slab [by changing $f_{\mathrm{PML}}=1+3i$ in Eq. (33) for solving PML-RQNMs$^{+}$ to $f_{\mathrm{PML}}=1-3i$ for solving PML-RQNMs$^{-}$], then the QCT can predict the PML-RQNMs$^{-}$ of the coupled slabs, which will include the physical QNMs with a negative real part of eigenfrequency.

### 3. Results of the QCT in predicting the source-excited scattered field of the two coupled slabs

As introduced in Sec. II A, the scattered field $\mathbf{\Psi}^{\mathrm{sca}}$ of the coupled system can be calculated by the QCT with Eq. (11) or (19) upon the basis of QNMs of each resonator or the coupled system, which are called Model 1 and Model 2 here, respectively. The accuracy of the two models will be tested numerically in this subsection.

The scattered field of the two coupled slabs excited by an $x$-polarized plane wave at frequency $\omega=\mathrm{Re}(\tilde{\omega}_{+,1})$ is shown in Fig. 6(a). By using all the solved PML-RQNMs (mode number $M$=802) of a single slab as a virtually complete basis, the results of Model 1 (blue solid lines) and Model 2 (black pluses) both agree with the analytical exact solution [red dot lines, given by Eq. (31)].

By using $M$=802 ESC-RQNMs [with the minimum $|\mathrm{Re}(\omega_{p,m})|$] of a single slab as a virtually complete basis, the results of Model 1 (green circles) also agree with the analytical exact solution. However, as stated in Sec. II B [after Eq. (28)], Model 2 using ESC-RQNMs is difficult to implement because it is difficult to find all the solutions of the exponentially nonlinear eigenvalue problem so as to obtain all the QNMs of the coupled system.

To quantitatively characterize the accuracy of QCT using RQNMs, a relative error $\sigma$ of scattered electric field $\mathbf{E}^{\mathrm{sca}}$ is defined as $\sigma=\left\|E_{x,\mathrm{QCT}}^{\mathrm{sca}}-E_{x,\mathrm{exact}}^{\mathrm{sca}}\right\|/\left\|E_{x,\mathrm{exact}}^{\mathrm{sca}}\right\|$, where $\left\|F(z)\right\|=\sqrt{\int_{-1.25d}^{1.25d}\left|F(z)\right|^{2}dz}$ represents the $L^2$-norm of function $F$, $E_{x,\mathrm{QCT}}^{\mathrm{sca}}$ and $E_{x,\mathrm{exact}}^{\mathrm{sca}}$ represent the electric fields obtained with the QCT (Model 1 or Model 2) and analytical exact solution, respectively. Figure 6(b) shows that for different excitation frequencies $\omega = \mathrm{Re}(\tilde{\omega}_{\pm,m})$ with $m$=1, 2, …, 12, the relative error $\sigma$ can be below $3\times10^{-3}$ for QCT using all the solved $M$=802 PML-RQNMs (blue solid line or black pluses), or below $2\times10^{-2}$ for QCT using $M$=802 ESC-RQNMs with the minimum $|\mathrm{Re}(\omega_{p,m})|$ (green solid line with circles). And for the QCT using all the solved PML-RQNMs, $\sigma$ is virtually the same for Model 1 (blue solid line) and Model 2 (black pluses). As shown in Fig. S4(b) in SM Sec. S4 [31], the $\sigma$ of QCT using PML-RQNMs can be further decreased by increasing $\mathrm{Im}(f_{\mathrm{PML}})$ [see Eq. (33)] to enhance the decay of PML-RQNM field in the PMLs.

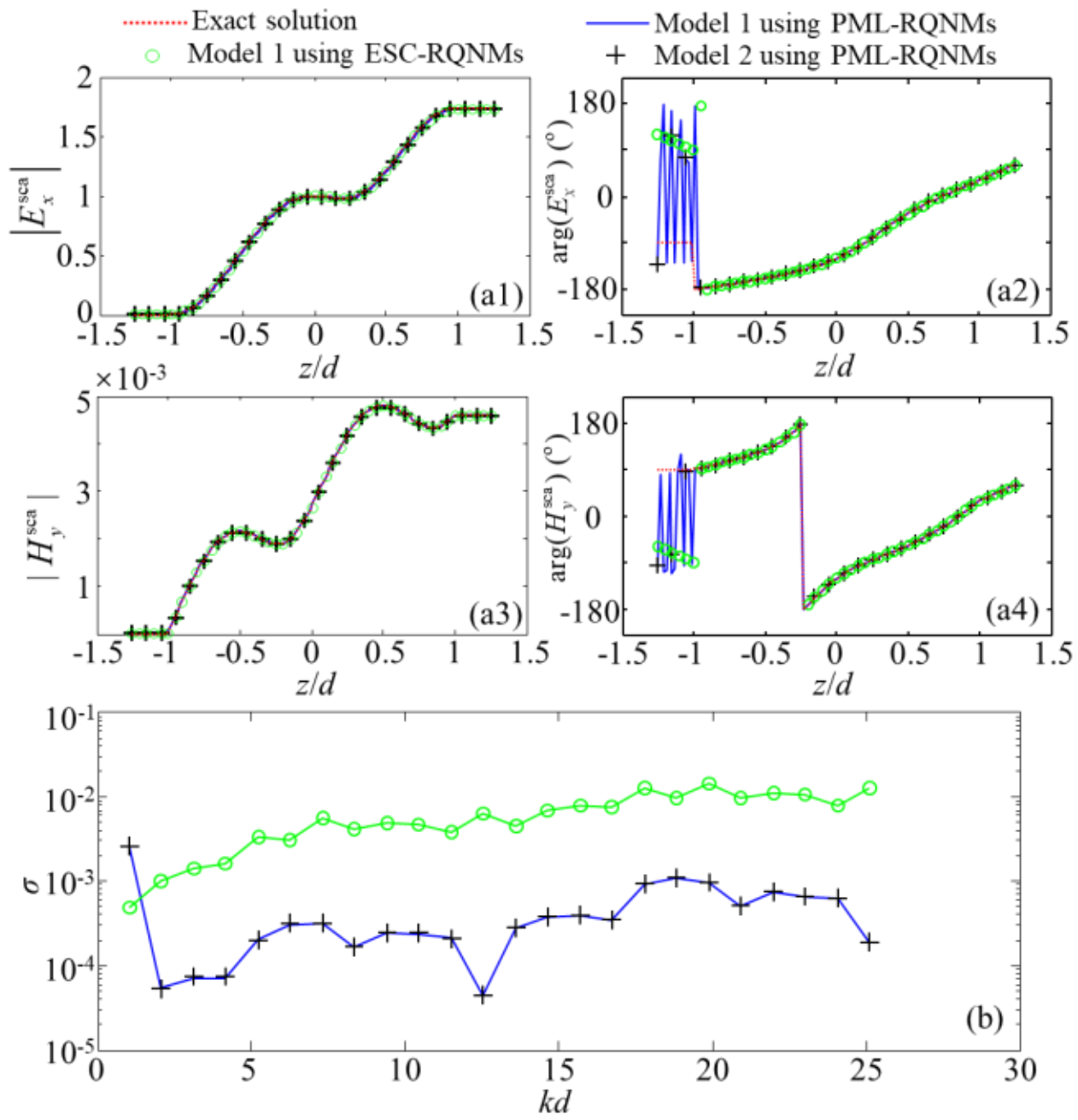

Fig. 6 (a) Scattered field of the two coupled slabs illuminated by a plane wave at frequency $\omega = \mathrm{Re}(\tilde{\omega}_{+,1})$ . The red dot lines are the results of analytical exact solution. The green circles show the predictions of the QCT Model 1 using $M$=802 ESC-RQNMs. The blue solid curves and black pluses show the predictions of the QCT Model 1 and Model 2 using all the solved $M$=802 PML-RQNMs of a single slab, respectively. (b) Relative error $\sigma$ of the QCT in predicting the scattered field at frequency $\omega = \mathrm{Re}(\tilde{\omega}_{\pm,m})$ ($m$=1, 2, …, 12). The curves in (b) follow the specifications in (a).

## IV. CONCLUSIONS

We report a strict numerical demonstration of the impact of mode completeness on the accuracy of the QCT established in Ref. [6], along with an improvement of the QCT and some theoretical demonstrations on a rigorous incorporation of the RQNMs into the QCT. The numerical example is selected as two coupled resonators of 1D slabs, for which a virtually complete basis of RQNMs can be solved either analytically (for ESC-RQNMs) or numerically (for PML-RQNMs). The extreme coupling case of direct contact of the two slabs is considered so as to provide a strong test of the accuracy of the QCT. The results show that by using a virtually complete basis of RQNMs, the QCT can achieve a high accuracy in predicting both the source-free eigenmodes and the source-excited scattered field of the coupled system, which however is not true if using the incomplete basis of not-regularized QNMs (i.e., physical QNMs). The present work provides a numerical evidence for the theoretically expected rigorousness of the QCT and the resultant ability for providing quantitative predictions.

## ACKNOWLEDGEMENT

The authors acknowledge helpful discussions with E. A. Muljarov at Cardiff University. This work was financially supported by the National Key Research and Development Program of China (2022YFA1404604) and the National Natural Science Foundation of China (12404434, 62475120, 12404438, 62535007).

[1] B. Peng, S. K. Özdemir, W. Chen, F. Nori, and L. Yang, What is and what is not electromagnetically induced transparency in whispering-gallery microcavities, Nat. Commun. **5**, 5082 (2014).
[2] J.-H. Park, A. Ndao, W. Cai, L. Hsu, A. Kodigala, T. Lepetit, Y.-H. Lo, and B. Kanté, Symmetry-breaking-induced plasmonic exceptional points and nanoscale sensing, Nat. Phys. **16**, 462 (2020).
[3] B. Gurlek, V. Sandoghdar, and D. Martín-Cano, Manipulation of quenching in nanoantenna–emitter systems enabled by external detuned cavities: A path to enhance strong-coupling, ACS Photonics **5**, 456 (2018).
[4] B. Vial and Y. Hao, A coupling model for quasi-normal modes of photonic resonators, J. Opt. **18**, 115004 (2016).
[5] P. T. Kristensen, K. Herrmann, F. Intravaia, and K. Busch, Modeling electromagnetic resonators using quasinormal modes, Adv. Opt. Photonics **12**, 612 (2020).
[6] C. Tao, J. D. Zhu, Y. Zhong, and H. Liu, Coupling theory of quasinormal modes for lossy and dispersive plasmonic nanoresonators, Phys. Rev. B **102**, 045430 (2020).
[7] J. Lin, M. Qiu, X. Zhang, H. Guo, Q. Cai, S. Xiao, Q. He, and L. Zhou, Tailoring the lineshapes of coupled plasmonic systems based on a theory derived from first principles, Light-Sci. Appl. **9**, 158 (2020).

[8] H. A. Haus and W. P. Huang, Coupled-mode theory, Proc. IEEE **79**, 1505 (1991).
[9] G. Benjamin, B. Jérémy, M. J. F. Olivier Martin, Numerical methods for nanophotonics: Standard problems and future challenges, Laser Photonics Rev. **9**, 577 (2015).
[10] P. T. Leung, S. Y. Liu, and K. Young, Completeness and orthogonality of quasinormal modes in leaky optical cavities, Phys. Rev. A **49**, 3057 (1994).
[11] E. A. Muljarov, W. Langbein, and R. Zimmermann, Brillouin-Wigner perturbation theory in open electromagnetic systems, EPL-Europhys. Lett. **92**, 50010 (2010).
[12] C. Sauvan, J. P. Hugonin, I. S. Maksymov, and P. Lalanne, Theory of the spontaneous optical emission of nanosize photonic and plasmon resonators, Phys. Rev. Lett. **110**, 237401 (2013).
[13] P. Lalanne, W. Yan, K. Vynck, C. Sauvan and J. Hugonin, Light interaction with photonic and plasmonic resonances, Laser Photonics Rev. **12**, 1700113 (2018).
[14] W. Yan, R. Faggiani, and P. Lalanne, Rigorous modal analysis of plasmonic nanoresonators, Phys. Rev. B **97**, 205422 (2018).
[15] C. Sauvan, T. Wu, R. Zarouf, E. A. Muljarov, and P. Lalanne, Normalization, orthogonality, and completeness of quasinormal modes of open systems: The case of electromagnetism [Invited], Opt. Express **30**, 6846 (2022).
[16] J. Ren, S. Franke, and S. Hughes, Quasinormal modes, local density of states, and classical Purcell factors for coupled loss-Gain resonators, Phys. Rev. X **11**, 041020 (2021).
[17] E. A. Muljarov, Rigorous theory of coupled resonators, Phys. Rev. Lett. **136**, 023801 (2026).
[18] T. Wu, P. Lalanne, Dissipative coupling in photonic and plasmonic resonators, Adv. Photonics, **7**, 056011 (2025).
[19] P. T. Leung and K. M. Pang, Completeness and time-independent perturbation of morphology-dependent resonances in dielectric spheres, J. Opt. Soc. Am. B **13**, 805 (1996).
[20] Z. Sztranyovszky, W. Langbein, and E. A. Muljarov, Extending completeness of the eigenmodes of an open system beyond its boundary, for Green's function and scattering matrix calculations, Phys. Rev. Res. **7**, L012035 (2025).
[21] B. Vial, F. Zolla, A. Nicolet, and M. Commandré, Quasimodal expansion of electromagnetic fields in open two-dimensional structures, Phys. Rev. A **89** (2014).
[22] A. Gras, P. Lalanne, and M. Durufle, Nonuniqueness of the quasinormal mode expansion of electromagnetic Lorentz dispersive materials, J. Opt. Soc. Am. A **37**, 1219 (2020).
[23] M. K. Dezfouli, S. Hughes, Regularized quasinormal modes for plasmonic resonators and open cavities, Phys. Rev. B **97**, 115302 (2018).
[24] R.-C. Ge, P. T. Kristensen, J. F. Young, and S. Hughes, Quasinormal mode approach to modelling light-emission and propagation in nanoplasmonics, New J. Phys. **16**, 113048 (2014).
[25] Q. Tao, Y. Su, Y. Zhong, H. Liu, Efficient method for modeling large-scale arrays of optical nanoresonators based on the coupling theory of quasinormal mode, Opt. Express, **32**, 7171 (2024).
[26] H. Zhao and H. Liu, Efficient electromagnetic modeling method for large-scale arrays of optical nanoresonators on a metallic substrate based on the coupling theory of quasinormal mode, Opt. Express **34**, 16802 (2026).
[27] E. Türeci, A. D. Stone, and B. Collier, Self-consistent multimode lasing theory for complex or random lasing media, Phys. Rev. A **74**, 043822 (2006).
[28] M. B. Doost, W. Langbein, and E. A. Muljarov, Resonant-state expansion applied to three-dimensional open optical systems, Phys. Rev. A **90**, 013834 (2014).
[29] Z. Qi, Y. Zhong, and H. Liu, Efficient method for the calculation of the optical force of multiple nanoparticles based on the coupling theory of quasinormal modes, Opt. Lett. **46**, 4610 (2021).
[30] E. A. Muljarov and T. Weiss, Resonant-state expansion for open optical systems: generalization to magnetic, chiral, and bi-anisotropic materials, Opt. Lett. **43**, 1978 (2018).
[31] See Supplemental Material at XXX, which includes details for the method for calculating the pseudoenergy $F_{p,m}$ of the QNM of a single resonator (Sec. S1), the plane-wave expansion method for solving the PML-RQNMs (Sec. S2), the spectral symmetry of PML-RQNMs of frequency-nondispersive PMLs and the legitimacy of using such PML-RQNMs for constructing the electromagnetic field in time domain (Sec. S3), the influence of the PML parameters on the accuracy of the QCT using PML-RQNMs (Sec. S4), and the results of the QCT using PML-RQNM$^-$ of a single slab in predicting the PML-RQNM$^-$ of the two coupled slabs (Sec. S5). References [5,12,14,15,22,30,32-35] are cited in the Supplemental Material.
[32] P. Lalanne, Effective properties and band structures of lamellar subwavelength crystals: Plane-wave method revisited, Phy. Rev. B **58**, 9801 (1998).
[33] J. P. Hugonin and P. Lalanne, Perfectly matched layers as nonlinear coordinate transforms: A generalized formalization, J. Opt. Soc. Am. A **22**, 1844 (2005).
[34] L. Li, Use of Fourier series in the analysis of discontinuous periodic structures, J. Opt. Soc. Am. A **13**, 1870 (1996).

[35] G. Demésy, T. Wu, Y. Brûlé, F. Zolla, A. Nicolet, P. Lalanne, and B. Gralak, Dispersive perfectly matched layers and high-order absorbing boundary conditions for electromagnetic quasinormal modes, J. Opt. Soc. Am. A **40**, 1947 (2023).

[36] H. S. Sehmi, W. Langbein, and E. A. Muljarov, Optimizing the Drude-Lorentz model for material permittivity: Method, program, and examples for gold, silver, and copper, Phys. Rev. B **95**, 115444 (2017).

[37] E. A. Muljarov and W. Langbein, Resonant-state expansion of dispersive open optical systems: Creating gold from sand, Phys. Rev. B **93**, 075417 (2016).

[38] J.-P. Dedieu and F. Tisseur, Perturbation theory for homogeneous polynomial eigenvalue problems, Linear Alg. Appl. **358**, 71-94 (2003).

[39] G. B. Arfken, H. J. Weber, and F. E. Harris, *Mathematical Methods for Physicists*, 6th ed. (Academic, Boston, 2005).

[40] C. A. Balanis, *Antenna Theory Analysis and Design*, 3rd ed. (John Wiley & Sons, Inc., 2005).

[41] J. Van Bladel, *Singular Electromagnetic Fields and Sources* (IEEE, New Jersey, 1991).

[42] According to the statement in Sec. I (paragraph 2), for resonators of analytically-solvable simple geometries [10,19] (such as the present numerical example of 1D slabs), the *Assumption 1* is true if the QNMs are chosen as the analytically-solvable physical QNMs and the $S_p$ is chosen as the boundary of the interior of the resonator. For resonators of general geometries, the *Assumption 1* is true if the QNMs are chosen as PML-RQNMs [14,15,21,22] (or physical QNMs along with the virtual gap modes generated by the resonant-state expansion [17,20]) that are numerically complete in expanding the scattered field $\mathbf{\Psi}_p^{\text{sca}}(\mathbf{r})$ inside and on the $S_p$ [i.e. $\mathbf{r}\in I(S_p)$], where there is no upper limit on the size of the $S_p$ but a larger $S_p$ requires solving more modes and resultantly, larger computational amount. Here note that the *Assumption 1* is actually a logical inference of Eq. (10) in Ref. [17], as explained below. For the scattered field $\mathbf{\Psi}_p^{\text{sca}}$, as shown by Eq. (3), the corresponding excitation source is an equivalent electric-current density,

$$\mathbf{J}_{\text{eq},p} = -i\omega\Delta\boldsymbol{\varepsilon}_p\cdot\left(\sum_{q\neq p}\mathbf{E}_q^{\text{sca}}+\mathbf{E}^{\text{b}}\right),$$

whose nonzero region is the nonzero region of the permittivity change $\Delta\boldsymbol{\varepsilon}_p$ (i.e., the region $V_p$ of the $p$th resonator) and therefore is contained by the $I(S_p)$, i.e. $V_p\subseteq I(S_p)$. This implies that if choosing $I(S_p)=V_j$ with $V_j$ being the notation in Eq. (10) of Ref. [17], the source position satisfies $\mathbf{r}'\in V_p\subseteq I(S_p)=V_j$, i.e. $\mathbf{r}'\in V_j$. This, along with the *Assumption 1*'s specification that the field position $\mathbf{r}\in I(S_p)=V_j$ for the scattered field $\mathbf{\Psi}_p^{\text{sca}}(\mathbf{r})$, fulfills the premise of Eq. (10) in Ref. [17], which then leads to the *Assumption 1*.

[43] Rigorously speaking, Eq. (28) will be different from Eq. (8) if and only if $V_p\cap O(S_q)\neq\varnothing$, which will lead to $\mathbf{E}_{q,n}^{\text{ESC}}(\mathbf{r},\omega)\neq\mathbf{E}_{q,n}(\mathbf{r}),\ \exists\mathbf{r}\in V_p\cap O(S_q)$. This implies that if $V_p\cap O(S_q)=\varnothing$ [i.e. $V_p\subseteq I(S_q)$], then there is,

$$\mathbf{E}_{q,n}^{\text{ESC}}(\mathbf{r},\omega)=\mathbf{E}_{q,n}(\mathbf{r}),\ \forall\mathbf{r}\in V_p\subseteq I(S_q),$$

so that Eq. (28) becomes the same as Eq. (8). This case can indeed occur, for instance, if the complete set of QNMs in the *Assumption 1* are chosen as a set of PML-RQNMs, or a set of modes including physical QNMs and virtual gap modes [20], for which the $I(S_q)$ is allowed to be distinctly larger than the $V_q$ or even out of the near-field region of the $q$th resonator [42], so that the case of $V_p\subseteq I(S_q)$ can occur.

[44] Y. Li, H. Liu, H. Jia, F. Bo, G. Zhang, and J. Xu, Fully vectorial modeling of cylindrical microresonators with aperiodic Fourier modal method, J. Opt. Soc. Am. A **31**, 2459 (2014).

[45] P. Lalanne et al., Quasinormal mode solvers for resonators with dispersive materials, J. Opt. Soc. Am. A **36**, 686 (2019).

## Supplemental Material

# Impact of mode completeness on the accuracy of the coupling theory of quasinormal modes: a strict numerical demonstration

Can Tao,[1,2,3] Junda Zhu[4], and Haitao Liu[1,2,*]

[1]*Institute of Modern Optics, College of Electronic Information and Optical Engineering, Nankai University, Tianjin, 300350, China*

[2]*National Key Laboratory of Semiconductor Laser, Tianjin, 300350, China*

[3]*The MOE Key Laboratory of Weak Light Nonlinear Photonics, TEDA Institute of Applied Physics and School of Physics, Nankai University, Tianjin 300457, China*

[4]*College of Physics and Materials Science, Tianjin Normal University, Tianjin 300387, China*

[*]Corresponding author: liuht@nankai.edu.cn

**CONTENTS**

### S1. Method for calculating the pseudoenergy $F_{p,m}$ of the QNM of a single resonator

The pseudoenergy $F_{p,m}$ of the $m$th-order QNM field $\mathbf{\Psi}_{p,m}$=[$\mathbf{E}_{p,m}$, $\mathbf{H}_{p,m}$]$^{\mathrm{T}}$ of the $p$th resonator is expressed as [1],

$$\begin{aligned}F_{p,m} = &\iiint_{\Omega}\{\mathbf{E}_{p,m}(\mathbf{r})\cdot\frac{\partial[\omega\boldsymbol{\varepsilon}_p(\mathbf{r},\omega)]}{\partial\omega}\cdot\mathbf{E}_{p,m}(\mathbf{r})-\mathbf{H}_{p,m}(\mathbf{r})\cdot\frac{\partial[\omega\boldsymbol{\mu}_p(\mathbf{r},\omega)]}{\partial\omega}\cdot\mathbf{H}_{p,m}(\mathbf{r})\}_{\omega=\omega_{p,m}}d^3\mathbf{r}\\ &+i\oiint_{\partial\Omega}[\mathbf{E}_{p,m}(\mathbf{r})\times\frac{\partial\mathbf{H}^S_{p,m}(\mathbf{r},\omega)}{\partial\omega}-\frac{\partial\mathbf{E}^S_{p,m}(\mathbf{r},\omega)}{\partial\omega}\times\mathbf{H}_{p,m}(\mathbf{r})]_{\omega=\omega_{p,m}}\cdot\mathbf{n}ds,\end{aligned}\tag{S1.1}$$

where $\boldsymbol{\varepsilon}_p$ and $\boldsymbol{\mu}_p$ are the permittivity and permeability tensors with the presence of only the $p$th scatterer in the background medium, $[\mathbf{E}^S_{p,m},\mathbf{H}^S_{p,m}]^{\mathrm{T}}$ is the electromagnetic field excited by an electric or magnetic current source with the density in the form of $\mathbf{J}(\mathbf{r},\omega)=(\omega-\omega_{p,m})\mathbf{S}(\mathbf{r})$ where $\mathbf{S}(\mathbf{r})$ is independent of the excitation frequency $\omega$, $\Omega$ is a region containing the current source, and $\mathbf{n}$ is the out-pointing unit normal vector on the boundary $\partial\Omega$ of $\Omega$. There is $[\mathbf{E}^S_{p,m}(\mathbf{r},\omega),\mathbf{H}^S_{p,m}(\mathbf{r},\omega)]^{\mathrm{T}}\rightarrow[\mathbf{E}_{p,m}(\mathbf{r}), \mathbf{H}_{p,m}(\mathbf{r})]^{\mathrm{T}}$ in the limit $\omega\rightarrow\omega_{p,m}$. For the PML-RQNMs, there is $\Omega=\Omega_{\mathrm{physical}}\cup\Omega_{\mathrm{PML}}$, where $\Omega_{\mathrm{PML}}$ and $\Omega_{\mathrm{physical}}$ represent the PML domain and the computational domain surrounded by the PMLs, respectively (see Sec. S3 B for more details). Then the surface integral in Eq. (S1.1) vanishes at the outer boundary of the PML, which leads to,

$$F_{p,m} = \iiint_{\Omega_{\text{physical}} \cup \Omega_{\text{PML}}} \{\mathbf{E}_{p,m}(\mathbf{r}) \cdot \frac{\partial[\omega \boldsymbol{\varepsilon}_p(\mathbf{r},\omega)]}{\partial \omega} \cdot \mathbf{E}_{p,m}(\mathbf{r}) - \mathbf{H}_{p,m}(\mathbf{r}) \cdot \frac{\partial[\omega \boldsymbol{\mu}_p(\mathbf{r},\omega)]}{\partial \omega} \cdot \mathbf{H}_{p,m}(\mathbf{r})\}_{\omega=\omega_{p,m}} d^3\mathbf{r}. \quad \text{(S1.2)}$$

For the one-dimensional (1D) slab made of nonmagnetic, isotropic and nondispersive material studied in the main text, the analytical expression for the electromagnetic field $\boldsymbol{\Psi}_{\pm,m}(z)=[\mathbf{E}_{\pm,m}, \mathbf{H}_{\pm,m}]^{\mathrm{T}}$ of the $m$th-order symmetric (+) or anti-symmetric (–) QNM is given by Eq. (30) in the main text. By substituting the expression of $\boldsymbol{\Psi}_{\pm,m}(z)$ into Eq. (S1.1), the second term of surface integral vanishes [2], and the remaining first term of volume integral gives the pseudoenergy $F_{p,\pm,m}=4(n_2)^2 d/r_{21}$.

## S2. Plane-wave expansion method for solving the PML-RQNMs

In the main text, the plane-wave expansion method (PWEM) is proposed for solving all the PML-RQNMs of the 1D slab, and details of the PWEM will be provided in this section. For the 1D slab made of nonmagnetic and isotropic medium, the electromagnetic field $\boldsymbol{\Psi}(z) = [\hat{\mathbf{x}}E_x(z), \hat{\mathbf{y}}H_y(z)]^{\mathrm{T}}$ of QNM satisfies the following source-free Maxwell's equations,

$$\frac{dH_y(z)}{dz} = ik\varepsilon_r(z)e_x(z), \ \frac{de_x(z)}{dz} = ik\mu_r(z)H_y(z), \quad \text{(S2.1)}$$

and satisfies the outgoing-wave condition at infinity ($z\rightarrow\pm\infty$). In Eq. (S2.1), $e_x=E_x/\eta_0$ is a normalized electric field ($\eta_0 = \sqrt{\mu_0/\varepsilon_0}$, $\varepsilon_0$ and $\mu_0$ being the wave impedance, permittivity and permeability in the vacuum, respectively), $k = \omega/c = \omega\sqrt{\mu_0\varepsilon_0}$ is the wave number in the vacuum ($c$ being the light speed in the vacuum), $\varepsilon_r$ and $\mu_r$ represent the relative permittivity and relative permeability, respectively, and there is $\mu_r=\mu_0$ for the presently considered nonmagnetic medium.

In the PWEM for solving the PML-RQNMs, the slab is surrounded by the PMLs, as shown in Fig. 3(b) in the main text. The PWEM introduced here is simply an application of the method in Ref. [3] to the special case of 1D structures but further incorporates the PMLs. The PML can be performed as a complex coordinate transformation $Z=f(z')$ of the Maxwell's equations [4], where $Z$ is an extension of the real-valued $z$ coordinate to complex plane (denoted as $z\rightarrow Z$), and $z'$ is a real-valued numerical coordinate. The $Z=f(z')$ adopted in the calculation is given by Eq. (33) in the main text. With the coordinate transformation $Z=f(z')$, Eq. (S2.1) becomes,

$$\frac{dH_y}{dz'} = ik\frac{df}{dz'}\varepsilon_r e_x = ik\varepsilon_r' e_x, \ \frac{de_x}{dz'} = ik\frac{df}{dz'}\mu_r H_y = ik\mu_r' H_y, \quad \text{(S2.2)}$$

where we define $\varepsilon_r' = \frac{df}{dz'}\varepsilon_r$ and $\mu_r' = \frac{df}{dz'}\mu_r$, and $df/dz'$ takes a complex value with a positive imaginary part as required for attenuating the outgoing wave [5]. The consistency between the forms of Eqs. (S2.1) and (S2.2) implies that besides the first viewpoint of complex coordinate transformation, a second viewpoint for the PML is an effective medium with an effective permittivity $\varepsilon_r'$ and effective permeability $\mu_r'$.

The incorporation of PMLs enables an implementation of the outgoing-wave condition satisfied by the PML-RQNMs, and brings a benefit that the outgoing far field of QNMs is transformed from exponential divergence in physical space to exponential decay in the PMLs [6], so that the QNMs become PML-RQNMs. By properly setting the parameters of PMLs, the field of PML-RQNMs at the outer boundary of PMLs can decay to nearly null value, so that the infinite physical space is mapped to a finite computational domain and a surrounding PML domain (their overall range denoted by $z_0 \leqslant z' \leqslant z_0+\Lambda$). Then by constructing an artificial periodic structure with the computational domain and the PML domain as a period, the $\varepsilon_r'(z')$, $\mu_r'(z')$ and electromagnetic field become periodic functions and can be expressed as Fourier series (or physically, plane-wave expansion),

$$\varepsilon_r'(z')=\sum_{m=-M_z}^{M_z}\varepsilon_{r,m}'\exp\left(im\frac{2\pi}{\Lambda}z'\right),\quad \mu_r'(z')=\sum_{m=-M_z}^{M_z}\mu_{r,m}'\exp\left(im\frac{2\pi}{\Lambda}z'\right), \tag{S2.3}$$

$$H_y(z')=\sum_{m=-M_z}^{M_z}H_{y,m}\exp\left(im\frac{2\pi}{\Lambda}z'\right),\quad e_x(z')=\sum_{m=-M_z}^{M_z}e_{x,m}\exp\left(im\frac{2\pi}{\Lambda}z'\right), \tag{S2.4}$$

where $M_z$ is the truncated Fourier or harmonic order. In view that the electromagnetic field components $e_x$ and $H_y$ are all continuous functions with respect to $z'$, the Fourier series of $\varepsilon_r' e_x$ and $\mu_r' H_y$ in the right side of Eq. (S2.2) can be expressed through the Laurent's rule [7] as,

$$\varepsilon_r'(z')e_x(z')=\sum_{m=-M_z}^{M_z}\left(\sum_{n=-M_z}^{M_z}\varepsilon_{r,m-n}'e_{x,n}\right)\exp\left(im\frac{2\pi}{\Lambda}z'\right), \tag{S2.5a}$$

$$\mu_r'(z')H_y(z')=\sum_{m=-M_z}^{M_z}\left(\sum_{n=-M_z}^{M_z}\mu_{r,m-n}'H_{y,n}\right)\exp\left(im\frac{2\pi}{\Lambda}z'\right). \tag{S2.5b}$$

Substituting Eqs. (S2.4) and (S2.5) into Eq. (S2.2), one can obtain,

$$H_{y,m}m\frac{2\pi}{\Lambda}=k\sum_{n=-M_z}^{M_z}\varepsilon_{r,m-n}'e_{x,n},\quad e_{x,m}m\frac{2\pi}{\Lambda}=k\sum_{n=-M_z}^{M_z}\mu_{r,m-n}'H_{y,n}, \tag{S2.6}$$

which can be rewritten as a matrix form,

$$\begin{bmatrix}\mathbf{W} & \mathbf{0}\\ \mathbf{0} & \mathbf{W}\end{bmatrix}\begin{bmatrix}\mathbf{h}\\ \mathbf{e}\end{bmatrix}=k\begin{bmatrix}\mathbf{0} & \mathbf{P}\\ \mathbf{U} & \mathbf{0}\end{bmatrix}\begin{bmatrix}\mathbf{h}\\ \mathbf{e}\end{bmatrix}, \tag{S2.7}$$

or equivalently,

$$\begin{bmatrix}\mathbf{0} & \mathbf{P}\\ \mathbf{U} & \mathbf{0}\end{bmatrix}^{-1}\begin{bmatrix}\mathbf{W} & \mathbf{0}\\ \mathbf{0} & \mathbf{W}\end{bmatrix}\begin{bmatrix}\mathbf{h}\\ \mathbf{e}\end{bmatrix}=k\begin{bmatrix}\mathbf{h}\\ \mathbf{e}\end{bmatrix}, \tag{S2.8}$$

where the $m$th-row and $n$th-column element of the matrices **P** and **U** are $\varepsilon_{r,m-n}'$ and $\mu_{r,m-n}'$, respectively, the $m$th diagonal element of the diagonal matrix **W** is $m2\pi/\Lambda$, and the $m$th elements of column vectors **h** and **e** are $H_{y,m}$ and $e_{x,m}$, respectively. Equation (S2.8) is a matrix eigenvalue problem, whose $s$th eigenvalue $k=k_s=\omega_s/c$ is the eigenfrequency of the $s$th PML-RQNM, and the corresponding eigenvector [**h**;**e**]=[$\mathbf{h}_s$;$\mathbf{e}_s$] gives the electromagnetic field of the $s$th PML-RQNM by using Eq. (S2.4).

The matrix eigenvalue problem (S2.8) brings a benefit that all the eigenvectors can be solved, and can form a numerically complete basis in expanding any vector of the same length, or equivalently, all the PML-RQNMs with a finite number can be solved, and can form a numerically complete basis in expanding any discretized electromagnetic field within the computational domain (containing both the inside and the outside of the resonator) truncated by the PMLs [5,8,9].

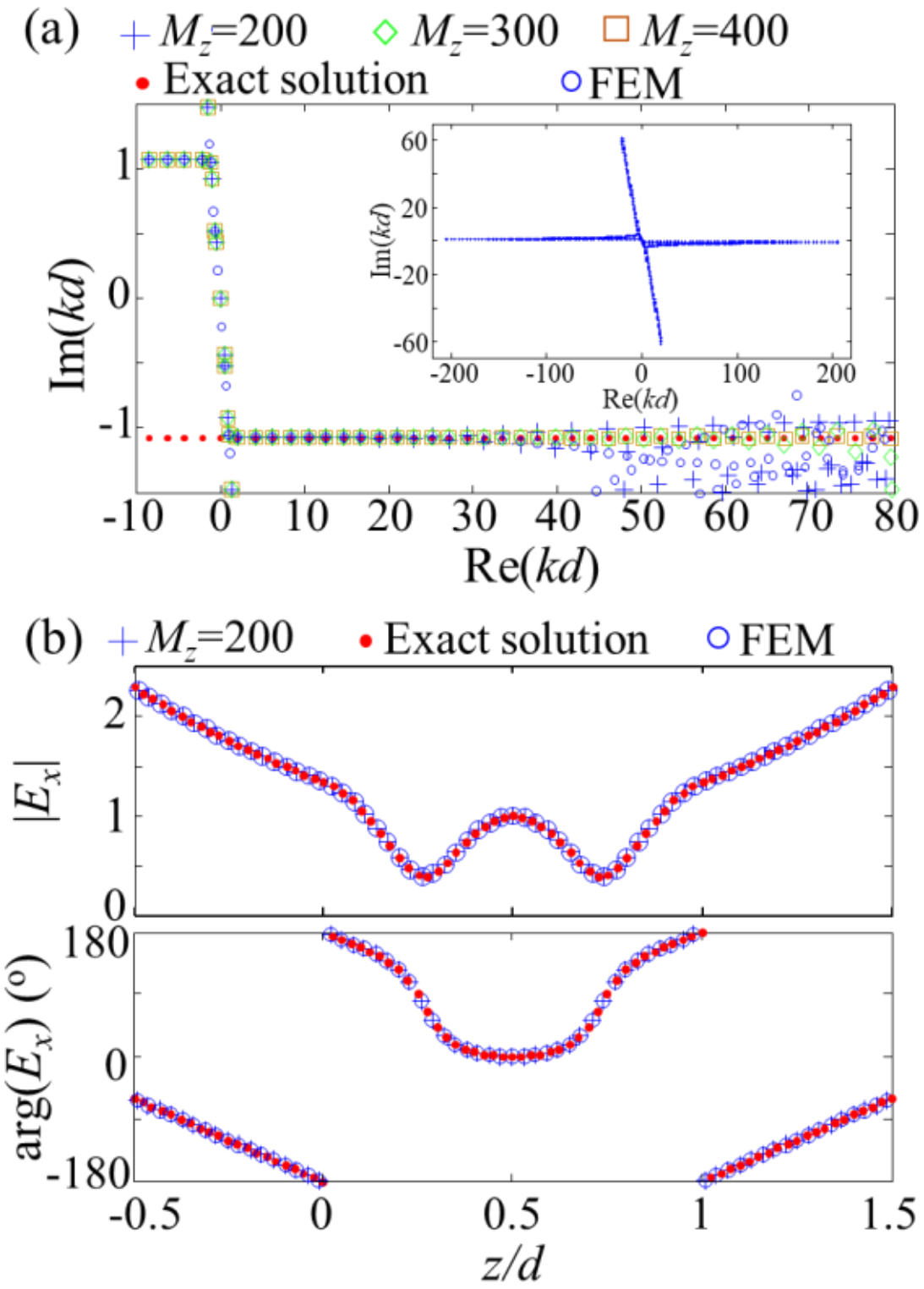

Fig. S1. Solution of QNMs of the single slab. (a) Eigenfrequencies of the solved QNMs. (b) The $x$-component of electric field of the 1st-order symmetric QNM (with eigenfrequency $\omega_{+,1}$) satisfying a normalization of $E_x(d/2)=1$, where the boundaries of the slab are at $z$=0 and $d$. In (a) and (b), the red dots are the results of the exact solutions (i.e., physical QNMs or not-regularized QNMs). The blue pluses and circles represent the numerical results of PML-RQNMs obtained with the PWEM (with $M_z$=200) and FEM, respectively. In (a), the green diamonds and brown squares are the results of PWEM with $M_z$=300 and 400, respectively. The inset in (a) shows a global distribution of the eigenfrequencies of all the PML-RQNMs solved with the PWEM (with $M_z$=200).

To test the validity of the above proposed PWEM, we compare the PML-RQNMs of the single slab solved by the PWEM and by the finite element method (FEM) which is implemented with the COMSOL Multiphysics software. In the PWEM, we set a truncated Fourier order $M_z$=200 and PML parameters consistent with those in the main text, i.e., $f_{\text{PML}}=1+3i$, $t_{\text{PML}}=d$, $z_L=z_1-1.5d$ and $z_R=z_2+1.5d$ [see Eq. (33) and the following setting of PML parameters in the main text], where $z_1$=0 and $z_2=d$ denote the boundary coordinates of the single slab. In the FEM, the degrees of freedom (i.e., number of unknowns) are 539, and the PML is implemented by setting an effective medium [as explained after Eq. (S2.2)] with parameters consistent with those of the PWEM. In the FEM, a perfect electric-conductor condition of $E_x$=0 is set at the two outer boundaries of the PMLs, which is different from the periodic condition adopted in the PWEM.

As discussed in Secs. II C and III B1 in the main text, the PML-RQNMs solved with the PWEM (and with the FEM as well) can be classified into physical and numerical PML-RQNMs, and the numerical PML-RQNMs can be further classified into transitional and purely numerical PML-RQNMs.

Figure S1(a) shows that for the physical PML-RQNMs solved with the PWEM (blue pluses), the eigenfrequencies agree well with the exact solution [red dots, given by Eq. (29) in the main text] and the corresponding results of FEM (blue circles), which confirms the validity of the proposed PWEM. For the numerical PML-RQNMs solved with the PWEM, the eigenfrequencies are different from the exact solution, and exhibit different but similar distribution compared with the FEM results. By increasing the truncated Fourier order $M_z$ of the PWEM (blue pluses, green diamonds and brown squares for $M_z$=200, 300 and 400, respectively), some numerical PML-RQNMs exhibit eigenfrequencies approaching the exact solution, and thus are classified into the transitional PML-RQNMs.

Figure S1(b) shows that for the 1st-order symmetric mode [with eigenfrequency $\omega_{+,1}$ given by Eq. (29a) in the main text], the electric field of the corresponding physical PML-RQNM solved with the PWEM (blue pluses) agrees well with the exact solution [red dots, given by Eq. (30) in the main text] and the corresponding result of FEM (blue circles), which again confirms the validity of the proposed PWEM.

**S3. Spectral symmetry of PML-RQNMs of frequency-nondispersive PMLs and legitimacy of using such PML-RQNMs for constructing the electromagnetic field in time domain**

For the 1D slab composed of frequency-nondispersive medium, Eq. (29) in the main text indicates that the eigenfrequencies of physical QNMs (i.e., not-regularized QNMs) satisfy a mirror symmetry with respect to the imaginary axis. However, this spectral symmetry is not satisfied by the PML-RQNMs, as shown in Fig. S1(a). In this section, we will explain that the spectral symmetry satisfied by the physical QNMs (proved in subsection A for the general case of frequency-dispersive medium) can be satisfied as well by the PML-RQNMs of frequency-nondispersive PMLs if considering all the PML-RQNMs for expanding both the positive-frequency and the negative-frequency components of electromagnetic field (theoretically demonstrated and numerically tested in subsection B), and that this spectral symmetry can ensure the legitimacy of using the PML-RQNMs of frequency-nondispersive PMLs for constructing the finally meaningful quantity of the electromagnetic field in time domain (demonstrated in subsection C). Besides, the central symmetry of eigenfrequencies with respect to the origin satisfied by the PML-RQNMs of frequency-nondispersive PMLs and medium [as shown in Fig. S1(a)] is also explained in subsection D.

**A. Spectral symmetry of physical QNMs (i.e., not-regularized QNMs)**

In this subsection, we will provide a theoretical demonstration of the spectral symmetry satisfied by the physical QNMs (i.e., not-regularized QNMs) for the general case of frequency-dispersive medium.

In the whole Sec. S3, the real-valued electromagnetic field in time domain is denoted by $\mathbf{\Psi}_t(t)=[\mathbf{E}_t(t),\mathbf{H}_t(t)]^{\mathrm{T}}$, and the complex-valued electromagnetic field in frequency domain is denoted by $\mathbf{\Psi}(\omega)=[\mathbf{E}(\omega),\mathbf{H}(\omega)]^{\mathrm{T}}$, where the dependence on spatial coordinates $\mathbf{r}=(x,y,z)$ is not written for simplicity, and will be written when necessary. $\mathbf{\Psi}(\omega)$ and $\mathbf{\Psi}_t(t)$ satisfy the Fourier transforms,

$$\mathbf{\Psi}(\omega)=\mathcal{F}_t^{-1}\{\mathbf{\Psi}_t(t)\}_v=\int_{-\infty}^{+\infty}\mathbf{\Psi}_t(t)\exp(i2\pi vt)dt, \tag{S3.1a}$$

$$\mathbf{\Psi}_t(t)=\mathcal{F}_v\{\mathbf{\Psi}(\omega)\}_t=\int_{-\infty}^{+\infty}\mathbf{\Psi}(\omega)\exp(-i2\pi vt)dv, \tag{S3.1b}$$

where $\nu=\omega/(2\pi)$ is the frequency with $\omega$ being the angular frequency. Note that logically, Eq. (S3.1a) is the definition of $\mathbf{\Psi}(\omega)$, and Eq. (S3.1b) is an inference from Eq. (S3.1a). $\mathbf{\Psi}(\omega)$ satisfies the frequency-domain Maxwell's equations,

$$\nabla\times\mathbf{E}(\omega)=i\omega\boldsymbol{\mu}(\omega)\cdot\mathbf{H}(\omega)-\mathbf{J}_{\rm m}(\omega),\quad \nabla\times\mathbf{H}(\omega)=-i\omega\boldsymbol{\varepsilon}(\omega)\cdot\mathbf{E}(\omega)+\mathbf{J}_{\rm e}(\omega), \tag{S3.2}$$

where $\boldsymbol{\varepsilon}(\omega)$ and $\boldsymbol{\mu}(\omega)$ are generally $\omega$-dependent permittivity and permeability tensors, respectively, and $\mathbf{J}_{\rm e}(\omega)$ and $\mathbf{J}_{\rm m}(\omega)$ are the electric and magnetic current density in frequency domain, respectively. The $\boldsymbol{\varepsilon}(\omega)$, $\boldsymbol{\mu}(\omega)$, $\mathbf{J}_{\rm m/e}(\omega)$ and $\mathbf{\Psi}(\omega)$ satisfy,

$$\boldsymbol{\varepsilon}(-\omega)=\boldsymbol{\varepsilon}(\omega)^*,\ \boldsymbol{\mu}(-\omega)=\boldsymbol{\mu}(\omega)^*,\ \mathbf{J}_{\rm m/e}(-\omega)=\mathbf{J}_{\rm m/e}(\omega)^*,\ \mathbf{\Psi}(-\omega)=\mathbf{\Psi}(\omega)^*,\ \omega\in\mathbb{R}, \tag{S3.3a}$$

$$\boldsymbol{\varepsilon}(-\omega^*)=\boldsymbol{\varepsilon}(\omega)^*,\ \boldsymbol{\mu}(-\omega^*)=\boldsymbol{\mu}(\omega)^*,\ \mathbf{J}_{\rm m/e}(-\omega^*)=\mathbf{J}_{\rm m/e}(\omega)^*,\ \mathbf{\Psi}(-\omega^*)=\mathbf{\Psi}(\omega)^*,\ \omega\in\mathbb{C}, \tag{S3.3b}$$

which can be derived from the fact that $\boldsymbol{\varepsilon}(\omega)$, $\boldsymbol{\mu}(\omega)$, $\mathbf{J}_{\rm m/e}(\omega)$ and $\mathbf{\Psi}(\omega)$ are all Fourier transforms of real-valued time-domain functions [see Eq. (S3.1a)].

According to the QNM-expansion theory [1,6,8], the electromagnetic field $\mathbf{\Psi}(\omega)$ excited by the source $\mathbf{J}_{\rm m/e}(\omega)$ can be expressed as,

$$\mathbf{\Psi}(\omega)=\sum_m\alpha_m(\omega)\mathbf{\Psi}(\omega_m), \tag{S3.4}$$

where $\mathbf{\Psi}(\omega_m)=[\mathbf{E}(\omega_m),\mathbf{H}(\omega_m)]^{\rm T}$ denotes the electromagnetic field of physical QNM (i.e., not-regularized QNM) at the complex eigenfrequency $\omega_m$ and satisfying source-free Maxwell's equations

$$\nabla\times\mathbf{E}(\omega_m)=i\omega_m\boldsymbol{\mu}(\omega_m)\cdot\mathbf{H}(\omega_m),\quad \nabla\times\mathbf{H}(\omega_m)=-i\omega_m\boldsymbol{\varepsilon}(\omega_m)\cdot\mathbf{E}(\omega_m). \tag{S3.5}$$

Note that the notation $\mathbf{\Psi}_{p,m}(\mathbf{r})$ of QNM used in the main text [see Eq. (4)] can be written as $\mathbf{\Psi}(\mathbf{r},\omega_{p,m})$ if using the notation here. Applying the complex conjugate to Eq. (S3.5), and then substituting the first two of Eq. (S3.3b) into it, one can obtain,

$$\nabla\times\mathbf{E}(\omega_m)^*=i(-\omega_m^*)\boldsymbol{\mu}(-\omega_m^*)\cdot\mathbf{H}(\omega_m)^*,\quad \nabla\times\mathbf{H}(\omega_m)^*=-i(-\omega_m^*)\boldsymbol{\varepsilon}(-\omega_m^*)\cdot\mathbf{E}(\omega_m)^*. \tag{S3.6}$$

Equations (S3.5) and (S3.6) imply that if there exists a QNM $\mathbf{\Psi}(\omega_m)$ with an eigenfrequency of $\omega_m$, then there must exist another QNM $\mathbf{\Psi}(-\omega_m^*)=[\mathbf{E}(-\omega_m^*),\mathbf{H}(-\omega_m^*)]^{\rm T}$ with an eigenfrequency of $-\omega_m^*$ and an electromagnetic field satisfying $\mathbf{E}(-\omega_m^*)=\mathbf{E}(\omega_m)^*,\ \mathbf{H}(-\omega_m^*)=\mathbf{H}(\omega_m)^*$, or written concisely as,

$$\mathbf{\Psi}(-\omega_m^*)=\mathbf{\Psi}(\omega_m)^*. \tag{S3.7}$$

In another word, Eq. (S3.7) means a QNM spectral symmetry that the eigenfrequencies must appear pairwise as $\omega_m$ and $-\omega_m^*$, i.e., a mirror symmetry with respect to the imaginary axis. Note that although Eq. (S3.7) and the last one of Eq. (S3.3b) have the same form, they are for essentially different cases without and with the presence of excitation source, respectively.

For the 1D slab composed of frequency-nondispersive medium studied in the main text, one can check that its physical QNMs [Eqs. (29) and (30)] satisfy the spectral symmetry of Eq. (S3.7). Besides, the spectral symmetry of Eq. (S3.7) is satisfied as well by the ESC-RQNMs according to their definition of Eq. (24) in the main text, i.e., $\mathbf{\Psi}^{\rm ESC}(\mathbf{r},-\omega_{p,m}^*,-\omega)=\mathbf{\Psi}^{\rm ESC}(\mathbf{r},\omega_{p,m},\omega)^*$, with $\mathbf{\Psi}^{\text{ESC}}(\mathbf{r},\omega_{p,m},\omega)$ denoting the $\mathbf{\Psi}_{p,m}^{\text{ESC}}(\mathbf{r},\omega)$ in Eq. (24).

**B. Spectral symmetry of PML-RQNMs of frequency-nondispersive PMLs**

As shown in Fig. S1(a), the PML-RQNMs of frequency-nondispersive PMLs do not satisfy the mirror symmetry of eigenfrequencies with respect to the imaginary axis, which appears to violate the spectral symmetry of Eq. (S3.7). However, as explained in this subsection, by considering all the PML-RQNMs for expanding both the positive-frequency and the negative-frequency components of electromagnetic field, which is required for constructing the electromagnetic field in time domain, the spectral symmetry of Eq. (S3.7) can be satisfied as well by the PML-RQNMs of frequency-nondispersive PMLs.

Because a finally meaningful use of QNMs is to construct the finally meaningful quantity of the electromagnetic field $\mathbf{\Psi}_t(t)$ in time domain, next we will try to find all the PML-RQNMs that are required to construct the $\mathbf{\Psi}_t(t)$. Equation (S3.1b) can be rewritten as,

$$\mathbf{\Psi}_t(t)=\int_{-\infty}^{0}\mathbf{\Psi}(\omega)\exp(-i2\pi vt)dv+\int_{0}^{+\infty}\mathbf{\Psi}(\omega)\exp(-i2\pi vt)dv, \tag{S3.8}$$

which shows that the construction of $\mathbf{\Psi}_t(t)$ requires both the *positive-frequency components* (the second term with $\omega>0$) and the *negative-frequency components* (the first term with $\omega<0$).

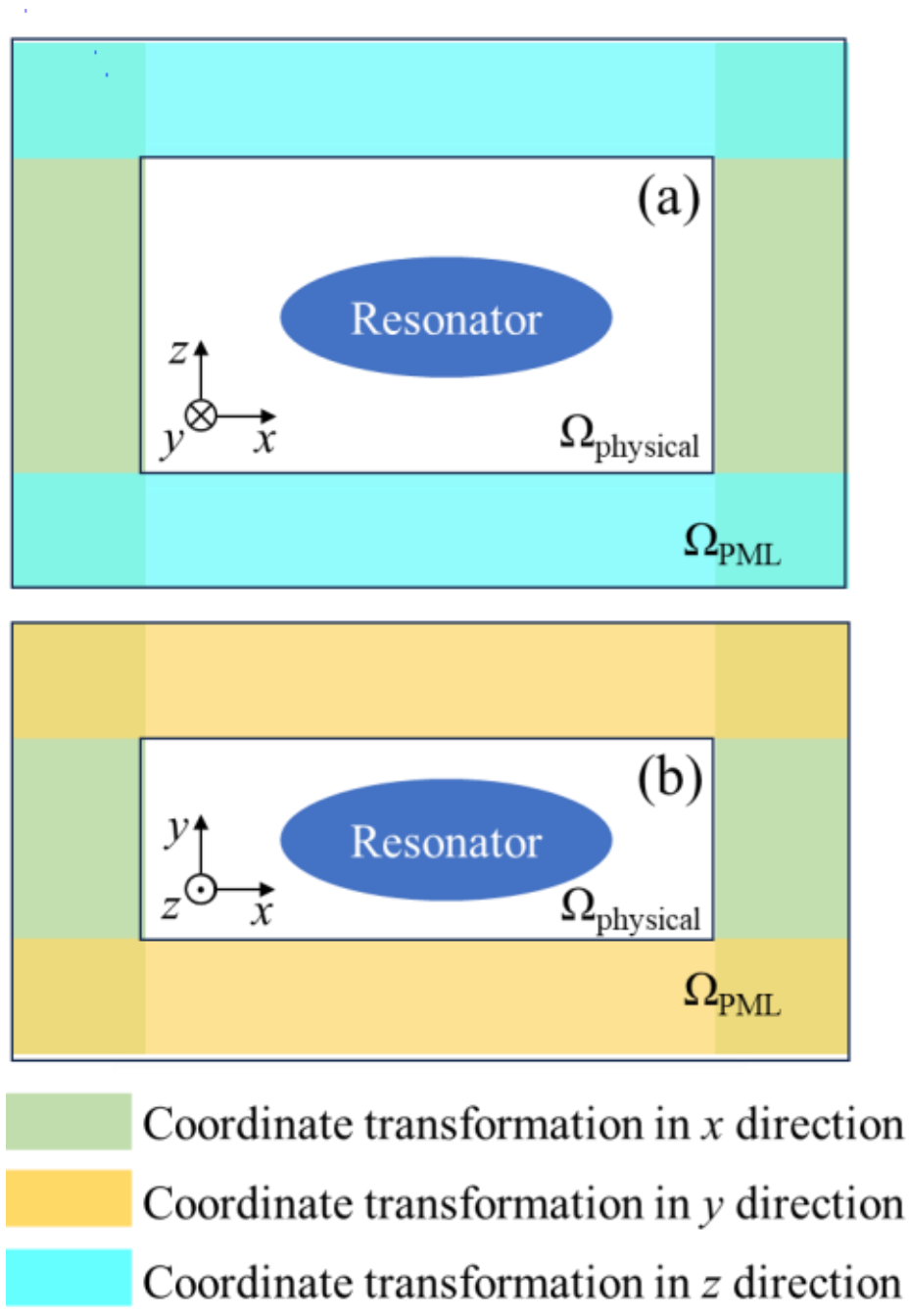

Fig. S2 Sketch of the PML domain $\Omega_{\text{PML}}$ and the computational domain $\Omega_{\text{physical}}$ surrounded by the PMLs for a three-dimensional resonator viewed in *x*-*z* plane (a) and *x*-*y* plane (b).

For the general case of PMLs along all the *x*, *y* and *z*-directions and anisotropic medium ($\boldsymbol{\varepsilon}$ and $\boldsymbol{\mu}$ being tensors) [6], analogous to Eq. (S2.2), the PMLs can be performed as a complex coordinate transformation (as illustrated in Fig. S2),

$$x\to X=f_X(x'),\ y\to Y=f_Y(y'),\ z\to Z=f_Z(z'), \tag{S3.9}$$

or equivalently, as an effective medium with an effective permittivity and permeability,

$$\boldsymbol{\varepsilon}'(\omega)=\mathbf{L}\cdot\boldsymbol{\varepsilon}(\omega)\cdot\mathbf{L}/\det(\mathbf{L}),\ \ \boldsymbol{\mu}'(\omega)=\mathbf{L}\cdot\boldsymbol{\mu}(\omega)\cdot\mathbf{L}/\det(\mathbf{L}), \tag{S3.10}$$

where $\mathbf{L}$ is a tensor defined as,

$$\mathbf{L}=(\hat{\mathbf{x}},\hat{\mathbf{y}},\hat{\mathbf{z}})\begin{bmatrix}(df_X/dx')^{-1} & & \\ & (df_Y/dy')^{-1} & \\ & & (df_Z/dz')^{-1}\end{bmatrix}\begin{bmatrix}\hat{\mathbf{x}}\\ \hat{\mathbf{y}}\\ \hat{\mathbf{z}}\end{bmatrix}, \tag{S3.11}$$

with $(\hat{\mathbf{x}},\hat{\mathbf{y}},\hat{\mathbf{z}})$ denoting the unit vectors along the axes of Cartesian ($x$, $y$, $z$) coordinates, respectively. In Eq. (S3.10), $\boldsymbol{\varepsilon}(\omega)$ and $\boldsymbol{\mu}(\omega)$ denote the permittivity and permeability tensors of the actual medium, respectively. In Eq. (S3.11), the ($f_X$, $f_Y$, $f_Z$) are defined such that for $(x, y, z)\in\Omega_{\text{PML}}$, the ($df_X/dx'$, $df_Y/dy'$, $df_Z/dz'$) all take complex value with a positive imaginary part and generally dependent on the frequency $\omega$ [10], and such that,

$$f_X(x')=x',\ f_Y(y')=y',\ f_Z(z')=z',\ \text{for } (x,y,z)\in\Omega_{\text{physical}}, \tag{S3.12}$$

where $\Omega_{\text{PML}}$ and $\Omega_{\text{physical}}$ represent the PML domain and the computational domain surrounded by the PMLs, respectively (as sketched in Fig. S2).

A frequency-nondispersive PML is defined such that the ($df_X/dx'$, $df_Y/dy'$, $df_Z/dz'$) are all independent of $\omega$ [they are possibly dependent on ($x'$, $y'$, $z'$)]. For instance, $df_Z/dz'=f_{\text{PML}}$ is a complex constant for the $f_Z(z')=f(z')$ defined in Eq. (33) in the main text. This definition of frequency-nondispersive PML will cause that the $\boldsymbol{\varepsilon}'(\omega)$ and $\boldsymbol{\mu}'(\omega)$ defined in Eq. (S3.10) violate the first two of Eq. (S3.3a), even if the $\boldsymbol{\varepsilon}(\omega)$ and $\boldsymbol{\mu}(\omega)$ satisfy the first two of Eq. (S3.3a). The key to reconcile this violation is that this setting of frequency-nondispersive PMLs [with the ($\mathbf{L}$, $\boldsymbol{\varepsilon}'(\omega)$, $\boldsymbol{\mu}'(\omega)$) denoted by ($\mathbf{L}^+$, $\boldsymbol{\varepsilon}'^+(\omega)$, $\boldsymbol{\mu}'^+(\omega)$)] is only for solving the *positive-frequency component* $\boldsymbol{\Psi}(\omega>0)$ and the corresponding PML-RQNMs (denoted by PML-RQNM$^+$) for expanding the $\boldsymbol{\Psi}(\omega>0)$. For solving the *negative-frequency component* $\boldsymbol{\Psi}(\omega<0)$ and the corresponding PML-RQNMs (denoted by PML-RQNM$^-$) for expanding the $\boldsymbol{\Psi}(\omega<0)$, the setting of frequency-nondispersive PMLs [with the ($\mathbf{L}$, $\boldsymbol{\varepsilon}'(\omega)$, $\boldsymbol{\mu}'(\omega)$) denoted by ($\mathbf{L}^-$, $\boldsymbol{\varepsilon}'^-(\omega)$, $\boldsymbol{\mu}'^-(\omega)$)] should satisfy,

$$\mathbf{L}^- = (\mathbf{L}^+)^*, \tag{S3.13a}$$

$$\boldsymbol{\varepsilon}'^-(-\omega)=\boldsymbol{\varepsilon}'^+(\omega)^*,\ \boldsymbol{\mu}'^-(-\omega)=\boldsymbol{\mu}'^+(\omega)^*,\ \omega>0, \tag{S3.13b}$$

$$\boldsymbol{\varepsilon}'^-(-\omega^*)=\boldsymbol{\varepsilon}'^+(\omega)^*,\ \boldsymbol{\mu}'^-(-\omega^*)=\boldsymbol{\mu}'^+(\omega)^*,\ \omega\in\mathbb{C}. \tag{S3.13c}$$

The above requirements lead to the following definitions,

$$\mathbf{L}(\omega)=\begin{cases}\mathbf{L}^+,\ \omega>0,\\ \mathbf{L}^-,\ \omega<0,\end{cases} \tag{S3.14a}$$

$$\boldsymbol{\varepsilon}'(\omega)=\left.\begin{cases}\boldsymbol{\varepsilon}'^+(\omega)=\mathbf{L}^+\cdot\boldsymbol{\varepsilon}^+(\omega)\cdot\mathbf{L}^+/\det(\mathbf{L}^+),\ \omega>0\\ \boldsymbol{\varepsilon}'^-(\omega)=\mathbf{L}^-\cdot\boldsymbol{\varepsilon}^-(\omega)\cdot\mathbf{L}^-/\det(\mathbf{L}^-),\ \omega<0\end{cases}\right\}=\mathbf{L}\cdot\boldsymbol{\varepsilon}(\omega)\cdot\mathbf{L}/\det(\mathbf{L}), \tag{S3.14b}$$

$$\boldsymbol{\mu}'(\omega)=\left.\begin{cases}\boldsymbol{\mu}'^+(\omega)=\mathbf{L}^+\cdot\boldsymbol{\mu}^+(\omega)\cdot\mathbf{L}^+/\det(\mathbf{L}^+),\ \omega>0\\ \boldsymbol{\mu}'^-(\omega)=\mathbf{L}^-\cdot\boldsymbol{\mu}^-(\omega)\cdot\mathbf{L}^-/\det(\mathbf{L}^-),\ \omega<0\end{cases}\right\}=\mathbf{L}\cdot\boldsymbol{\mu}(\omega)\cdot\mathbf{L}/\det(\mathbf{L}), \tag{S3.14c}$$

where the $\boldsymbol{\varepsilon}(\omega)$ and $\boldsymbol{\mu}(\omega)$ are defined as,

$$\boldsymbol{\varepsilon}(\omega)=\begin{cases}\boldsymbol{\varepsilon}^{+}(\omega), & \omega>0\\ \boldsymbol{\varepsilon}^{-}(\omega), & \omega<0\end{cases}\text{ (a)},\ \boldsymbol{\mu}(\omega)=\begin{cases}\boldsymbol{\mu}^{+}(\omega), & \omega>0\\ \boldsymbol{\mu}^{-}(\omega), & \omega<0\end{cases}\text{ (b)}. \quad \text{(S3.15)}$$

To fulfil Eqs. (S3.13b) and (S3.13c), the $\boldsymbol{\varepsilon}^{\pm}(\omega)$ and $\boldsymbol{\mu}^{\pm}(\omega)$ should satisfy,

$$\boldsymbol{\varepsilon}^{-}(-\omega)=\boldsymbol{\varepsilon}^{+}(\omega)^{*},\ \boldsymbol{\mu}^{-}(-\omega)=\boldsymbol{\mu}^{+}(\omega)^{*},\ \omega>0, \quad \text{(S3.16a)}$$

$$\boldsymbol{\varepsilon}^{-}(-\omega^{*})=\boldsymbol{\varepsilon}^{+}(\omega)^{*},\ \boldsymbol{\mu}^{-}(-\omega^{*})=\boldsymbol{\mu}^{+}(\omega)^{*},\ \omega\in\mathbb{C}. \quad \text{(S3.16b)}$$

Note that Eqs. (S3.15) and (S3.16) allow that the $\boldsymbol{\varepsilon}(\omega)$ and $\boldsymbol{\mu}(\omega)$ have different expressions for $\omega>0$ and $\omega<0$, which however is not allowed by the first two of Eq. (S3.3). Therefore, the former represents a more general case than the latter, and in an approximate manner, can describe the medium with a weak frequency dispersion excited by electromagnetic field with a narrow frequency range. For instance, for lossy and frequency-nondispersive medium, there is $\boldsymbol{\varepsilon}^{-}=(\boldsymbol{\varepsilon}^{+})^{*}\neq\boldsymbol{\varepsilon}^{+}$ or $\boldsymbol{\mu}^{-}=(\boldsymbol{\mu}^{+})^{*}\neq\boldsymbol{\mu}^{+}$ taking complex values independent of $\omega$, which can be described by Eqs. (S3.15) and (S3.16) but not by Eq. (S3.3).

By using Eq. (S3.14) and applying the complex coordinate transformation (S3.9) of PMLs to Eq. (S3.2) for $\omega>0$ and $\omega<0$, respectively, one can obtain that the $\boldsymbol{\Psi}(\omega>0)$ and $\boldsymbol{\Psi}(\omega<0)$ respectively satisfy the following effective-medium Maxwell's equations,

$$\nabla'\times\mathbf{E}'(\omega)=i\omega\boldsymbol{\mu}'^{+}(\omega)\cdot\mathbf{H}'(\omega)-\mathbf{J}_{\mathrm{m}}(\omega),\quad \nabla'\times\mathbf{H}'(\omega)=-i\omega\boldsymbol{\varepsilon}'^{+}(\omega)\cdot\mathbf{E}'(\omega)+\mathbf{J}_{\mathrm{e}}(\omega),\quad \omega>0, \quad \text{(S3.17a)}$$

$$\nabla'\times\mathbf{E}'(\omega)=i\omega\boldsymbol{\mu}'^{-}(\omega)\cdot\mathbf{H}'(\omega)-\mathbf{J}_{\mathrm{m}}(\omega),\quad \nabla'\times\mathbf{H}'(\omega)=-i\omega\boldsymbol{\varepsilon}'^{-}(\omega)\cdot\mathbf{E}'(\omega)+\mathbf{J}_{\mathrm{e}}(\omega),\quad \omega<0, \quad \text{(S3.17b)}$$

with

$$\nabla'=\hat{\mathbf{x}}\frac{\partial}{\partial x'}+\hat{\mathbf{y}}\frac{\partial}{\partial y'}+\hat{\mathbf{z}}\frac{\partial}{\partial z'}\text{ (a)},\ \mathbf{E}'(\omega)=\mathbf{L}^{-1}\cdot\mathbf{E}(\omega)\text{ (b)},\ \mathbf{H}'(\omega)=\mathbf{L}^{-1}\cdot\mathbf{H}(\omega)\text{ (c)}, \quad \text{(S3.18)}$$

where the differential operator $\nabla'$ performs derivatives with respect to the numerical coordinates ($x'$, $y'$, $z'$) [similar to Eq. (S2.2)], and $\mathbf{E}'(\omega)$ and $\mathbf{H}'(\omega)$ are defined as the effective electric- and magnetic-field vectors, respectively. Here note that due to Eq. (S3.12), there is,

$$\boldsymbol{\Psi}'(\mathbf{r},\omega)=\boldsymbol{\Psi}(\mathbf{r},\omega),\ \text{for } \mathbf{r}=(x,y,z)\in\Omega_{\text{physical}}, \quad \text{(S3.19)}$$

where $\boldsymbol{\Psi}=[\mathbf{E},\mathbf{H}]^{\mathrm{T}}$ and $\boldsymbol{\Psi}'=[\mathbf{E}',\mathbf{H}']^{\mathrm{T}}$. Equation (S3.17) can be collectively rewritten as,

$$\nabla'\times\mathbf{E}'(\omega)=i\omega\boldsymbol{\mu}'(\omega)\cdot\mathbf{H}'(\omega)-\mathbf{J}_{\mathrm{m}}(\omega),\quad \nabla'\times\mathbf{H}'(\omega)=-i\omega\boldsymbol{\varepsilon}'(\omega)\cdot\mathbf{E}'(\omega)+\mathbf{J}_{\mathrm{e}}(\omega),\quad \omega\in\mathbb{R}, \quad \text{(S3.20)}$$

which takes the same form as Eq. (S3.2). Similarly to Eq. (S3.2), one can check that the $\boldsymbol{\varepsilon}'(\omega)$ and $\boldsymbol{\mu}'(\omega)$ in Eq. (S3.20) really satisfy the first two of Eq. (S3.3a) by using Eqs. (S3.13b) and (S3.14). However, different from Eq. (S3.2), the $\boldsymbol{\varepsilon}'(\omega)$ and $\boldsymbol{\mu}'(\omega)$ in Eq. (S3.20) do not satisfy the first two of Eq. (S3.3b) even using Eq. (S3.13c), because an analytic continuation of $\boldsymbol{\varepsilon}'(\omega)$ and $\boldsymbol{\mu}'(\omega)$ from real axis to complex plane of frequency is prohibited by Eq. (S3.14a), where $\mathbf{L}^{+}$ and $\mathbf{L}^{-}$ generally take different complex values [satisfying Eq. (S3.13a)] independent of $\omega$ for frequency-nondispersive PMLs.

In accordance with Eq. (S3.17), the PML-RQNM$^{+}$ for expanding the $\boldsymbol{\Psi}(\omega>0)$ and the PML-RQNM$^{-}$ for expanding $\boldsymbol{\Psi}(\omega<0)$ [see the statement before Eq. (S3.13)] should respectively satisfy the following source-free Maxwell's equations,

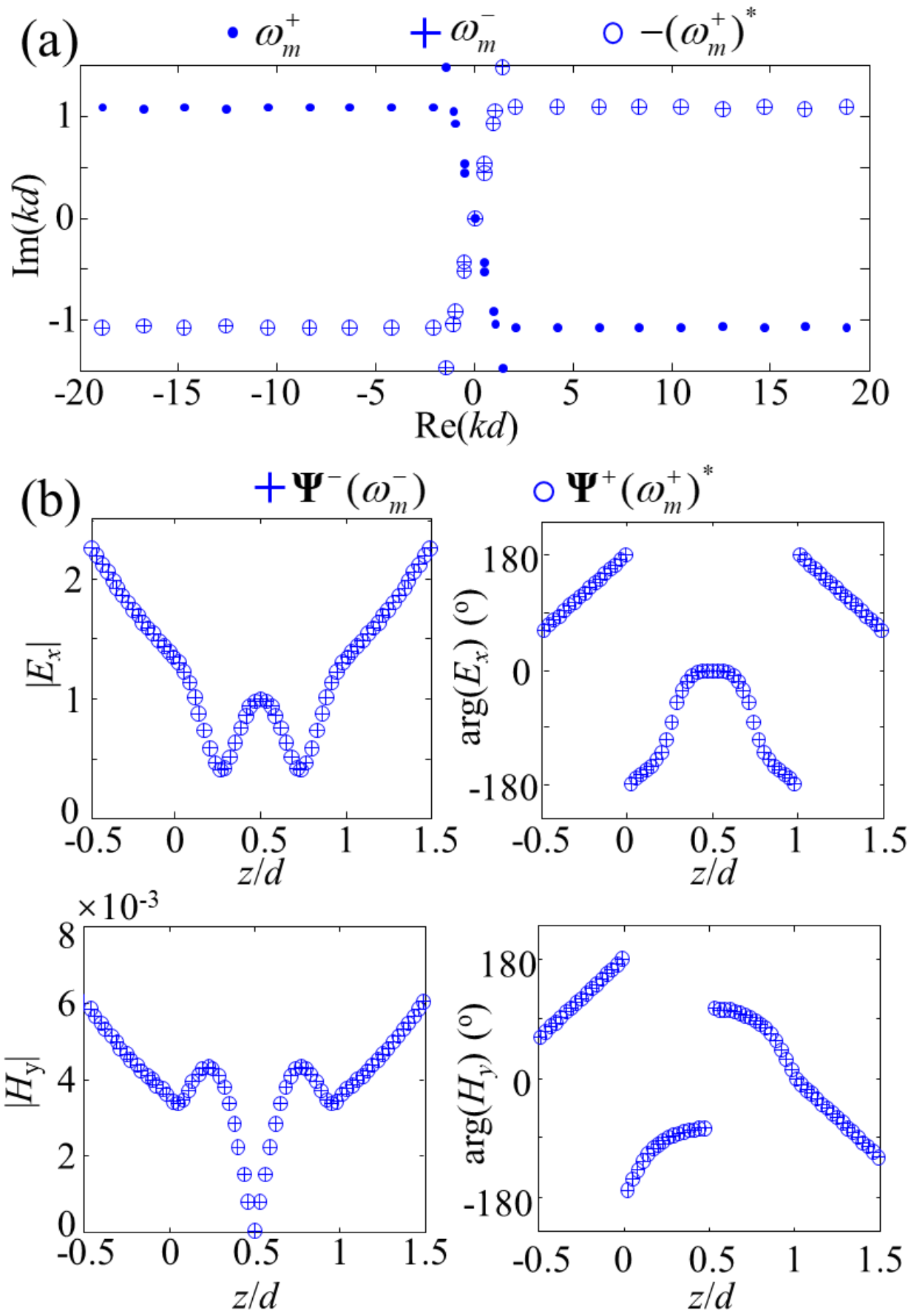

Fig. S3 (a) Eigenfrequencies $\omega_m^+$ (dots), $\omega_m^-$ (pluses), and $-(\omega_m^+)^*$ (circles) of the PML-RQNMs of the single slab. (b) Electromagnetic field $\mathbf{\Psi}^-(\omega_m^-)$ (pluses) and $\mathbf{\Psi}^+(\omega_m^+)^*$ (circles) of the PML-RQNM of the single slab, where there is $\omega_m^- = -(\omega_m^+)^*$ and $\mathbf{\Psi}^+(\omega_m^+)$ corresponds to the 1st-order symmetric physical QNM [with eigenfrequency $\omega_{+,1}$ in Eq. (29a) in the main text]. The results are obtained with the PWEM (truncated harmonic order $M_z$=200).

$$\nabla' \times \mathbf{E}'^+(\omega_m^+) = i\omega_m^+ \boldsymbol{\mu}'^+(\omega_m^+) \cdot \mathbf{H}'^+(\omega_m^+), \quad \nabla' \times \mathbf{H}'^+(\omega_m^+) = -i\omega_m^+ \boldsymbol{\varepsilon}'^+(\omega_m^+) \cdot \mathbf{E}'^+(\omega_m^+), \tag{S3.21a}$$

$$\nabla' \times \mathbf{E}'^-(\omega_m^-) = i\omega_m^- \boldsymbol{\mu}'^-(\omega_m^-) \cdot \mathbf{H}'^-(\omega_m^-), \quad \nabla' \times \mathbf{H}'^-(\omega_m^-) = -i\omega_m^- \boldsymbol{\varepsilon}'^-(\omega_m^-) \cdot \mathbf{E}'^-(\omega_m^-), \tag{S3.21b}$$

where $\mathbf{\Psi}'^{\pm}(\omega_m^{\pm}) = [\mathbf{E}'^{\pm}(\omega_m^{\pm}), \mathbf{H}'^{\pm}(\omega_m^{\pm})]^{\mathrm{T}}$ denotes the electromagnetic field of PML-RQNM$^{\pm}$ with eigenfrequency $\omega_m^{\pm}$. Similar to Eq. (S3.6), by applying the complex conjugate to Eq. (S3.21a), and then substituting Eq. (S3.13c) into it, one can obtain,

$$\nabla' \times \mathbf{E}'^+(\omega_m^+)^* = i[-(\omega_m^+)^*] \boldsymbol{\mu}'^-(-(\omega_m^+)^*) \cdot \mathbf{H}'^+(\omega_m^+)^*, \tag{S3.22a}$$

$$\nabla' \times \mathbf{H}'^+(\omega_m^+)^* = -i[-(\omega_m^+)^*] \boldsymbol{\varepsilon}'^-(-(\omega_m^+)^*) \cdot \mathbf{E}'^+(\omega_m^+)^*. \tag{S3.22b}$$

Comparing Eqs. (S3.21b) and (S3.22), one finally obtains,

$$\omega_m^- = -(\omega_m^+)^* \text{ (a)}, \quad \mathbf{\Psi}'^-(\omega_m^-) = \mathbf{\Psi}'^+(\omega_m^+)^* \text{ (b)}. \tag{S3.23}$$

Equation (S3.23b) can lead to,

$$\mathbf{\Psi}^+(\omega_m^+)^* = [\mathbf{L}^+ \cdot \mathbf{\Psi}'^+(\omega_m^+)]^* = (\mathbf{L}^+)^* \cdot \mathbf{\Psi}'^+(\omega_m^+)^* = \mathbf{L}^- \cdot \mathbf{\Psi}'^-(\omega_m^-) = \mathbf{\Psi}^-(\omega_m^-), \tag{S3.24}$$

where the first and last equalities are obtained by using Eqs. (S3.18b) and (S3.18c), and the third equality is obtained by using Eq. (S3.13a) and (S3.23b). Equation (S3.23) gives the spectral symmetry satisfied by the PML-RQNMs of frequency-nondispersive PMLs. Equation (S3.23) is consistent with Eq. (S3.7), which implies that the spectral symmetry of Eq. (S3.7) satisfied by the physical QNMs can be satisfied as well by the PML-RQNMs of frequency-nondispersive PMLs, provided that all the PML-RQNMs [i.e., PML-RQNM$^+$ $\mathbf{\Psi}^+(\omega_m^+)$ and PML-RQNM$^-$ $\mathbf{\Psi}^-(\omega_m^-)$ ] for expanding both the positive-frequency and the negative-frequency components of electromagnetic field are considered. Equation (S3.23) also shows that the PML-RQNM$^-$ can be obtained automatically from PML-RQNM$^+$ without solving the source-free Maxwell's equations (S3.21b).

As shown in Fig. S3, the eigenfrequencies and electromagnetic field of the PML-RQNMs of the single slab satisfy Eq. (S3.23), thus verifying its validity. Here for solving the PML-RQNMs$^-$ and PML-RQNMs$^+$ of the single slab with the full-wave PWEM, the corresponding frequency-nondispersive PMLs are formulated by setting $f_{\text{PML}}=1-3i$ and $1+3i$, respectively, and consistently setting $t_{\text{PML}}=d$, $z_L=z_1-1.5d$, and $z_R=z_2+1.5d$ [see their definitions in Eq. (33) in the main text], where $z_1=0$ and $z_2=d$ denote the boundary coordinates of the single slab. Such setting of PML parameters then ensures the satisfaction of Eq. (S3.13a) which is the premise of Eq. (S3.23). The truncated harmonic order is set to be $M_z=200$ for solving both the PML-RQNMs$^-$ and the PML-RQNMs$^+$. Note that the above PWEM parameters for solving the PML-RQNMs$^+$ are consistent with those adopted in the main text.

**C. Legitimacy of using PML-RQNMs of frequency-nondispersive PMLs for constructing the electromagnetic field in time domain**

A finally meaningful use of QNMs is to construct the finally meaningful quantity of the electromagnetic field $\mathbf{\Psi}_t(t)$ in time domain [8]. In this subsection, we will demonstrate the legitimacy of using all the PML-RQNMs [including PML-RQNM$^+$ $\mathbf{\Psi}^+(\omega_m^+)$ and PML-RQNM$^-$ $\mathbf{\Psi}^-(\omega_m^-)$ ] of frequency-nondispersive PMLs for constructing the real-valued $\mathbf{\Psi}_t(t)$.

According to the QNM-expansion theory [1,6,8], the positive-frequency component $\mathbf{\Psi}(\omega)$ and negative-frequency component $\mathbf{\Psi}(-\omega)$ with $\omega>0$ can be expressed as,

$$\mathbf{\Psi}(\pm\omega) = \sum_m \alpha_m^\pm(\pm\omega)\mathbf{\Psi}^\pm(\omega_m^\pm), \ \omega > 0, \tag{S3.25}$$

where $\mathbf{\Psi}^\pm(\omega_m^\pm) = [\mathbf{E}^\pm(\omega_m^\pm), \mathbf{H}^\pm(\omega_m^\pm)]^{\text{T}}$ denotes the electromagnetic field of the PML-RQNM$^\pm$ with eigenfrequency $\omega_m^\pm$, and the QNM-expansion coefficient is,

$$\alpha_m^\pm(\pm\omega) = \frac{\iiint_{R^3}[\mathbf{E}^\pm(\omega_m^\pm)\cdot\mathbf{J}_{\text{e}}(\pm\omega) - \mathbf{H}^\pm(\omega_m^\pm)\cdot\mathbf{J}_{\text{m}}(\pm\omega)]d^3\mathbf{r}}{i(\pm\omega-\omega_m^\pm)F_m^\pm}, \ \omega > 0, \tag{S3.26}$$

with the pseudoenergy [6],

$$F_m^{\pm} = \iiint_{\Omega_{\text{physical}} \cup \Omega_{\text{PML}}} \left\{ \mathbf{E}^{\pm}(\omega_m^{\pm}) \cdot \left.\frac{\partial[\omega \boldsymbol{\varepsilon}^{\pm}(\omega)]}{\partial \omega}\right|_{\omega=\omega_m^{\pm}} \cdot \mathbf{E}^{\pm}(\omega_m^{\pm}) - \mathbf{H}^{\pm}(\omega_m^{\pm}) \cdot \left.\frac{\partial[\omega \boldsymbol{\mu}^{\pm}(\omega)]}{\partial \omega}\right|_{\omega=\omega_m^{\pm}} \cdot \mathbf{H}^{\pm}(\omega_m^{\pm}) \right\} d^3\mathbf{r}. \quad \text{(S3.27)}$$

In the following, we will prove that the $\boldsymbol{\Psi}(\pm\omega)$ given by Eqs. (S3.25)-(S3.27) consistently satisfy the last one of Eq. (S3.3a), so that the $\boldsymbol{\Psi}_t(t)$ constructed by the $\boldsymbol{\Psi}(\pm\omega)$ [see Eq. (S3.1b)] consistently takes real values, which then ensures the legitimacy of using PML-RQNMs of frequency-nondispersive PMLs for constructing the $\boldsymbol{\Psi}_t(t)$.

*Proof*: Applying the complex conjugate to Eq. (S3.25) for $\boldsymbol{\Psi}(\omega>0)$, one can obtain,

$$\boldsymbol{\Psi}(\omega>0)^* = \sum_m \alpha_m^+(\omega)^* \boldsymbol{\Psi}^+(\omega_m^+)^* = \sum_m \alpha_m^-(-\omega) \boldsymbol{\Psi}^-(\omega_m^-) = \boldsymbol{\Psi}(-\omega). \quad \text{(S3.28)}$$

where the second equality is obtained by using Eq. (S3.24) and

$$\begin{aligned} \alpha_m^+(\omega)^* &= \frac{\iiint_{R^3} [\mathbf{E}^+(\omega_m^+)^* \cdot \mathbf{J}_{\mathrm{e}}(\omega)^* - \mathbf{H}^+(\omega_m^+)^* \cdot \mathbf{J}_{\mathrm{m}}(\omega)^*] d^3\mathbf{r}}{-i[\omega - (\omega_m^+)^*](F_m^+)^*} \\ &= \frac{\iiint_{R^3} [\mathbf{E}^-(\omega_m^-) \cdot \mathbf{J}_{\mathrm{e}}(-\omega) - \mathbf{H}^-(\omega_m^-) \cdot \mathbf{J}_{\mathrm{m}}(-\omega)] d^3\mathbf{r}}{i(-\omega - \omega_m^-) F_m^-} \\ &= \alpha_m^-(-\omega). \end{aligned} \quad \text{(S3.29)}$$

In Eq. (S3.29), the second equality is derived by using Eq. (S3.23a), (S3.24), the third one of Eq. (S3.3a), and

$$(F_m^+)^* = F_m^-, \quad \text{(S3.30)}$$

*which will be proved hereafter*. The last equality of Eq. (S3.28) simply gives the last one of Eq. (S3.3a). Substituting Eq. (S3.28) into Eq. (S3.1b), one finally obtains,

$$\begin{aligned} \boldsymbol{\Psi}_t(t) &= \int_{-\infty}^{0} \boldsymbol{\Psi}(\omega) \exp(-i2\pi v t) dv + \int_{0}^{+\infty} \boldsymbol{\Psi}(\omega) \exp(-i2\pi v t) dv \\ &= \int_{0}^{+\infty} \boldsymbol{\Psi}(-\omega') \exp(i2\pi v' t) dv' + \int_{0}^{+\infty} \boldsymbol{\Psi}(\omega) \exp(-i2\pi v t) dv \\ &= \int_{0}^{+\infty} \boldsymbol{\Psi}(\omega')^* \exp(i2\pi v' t) dv' + \int_{0}^{+\infty} \boldsymbol{\Psi}(\omega) \exp(-i2\pi v t) dv \\ &= 2\,\mathrm{Re}\left[ \int_{0}^{+\infty} \boldsymbol{\Psi}(\omega) \exp(-i2\pi v t) dv \right] \in \mathbb{R}, \end{aligned} \quad \text{(S3.31)}$$

where $\omega'=-\omega$ and $v'=\omega'/(2\pi)$. Equation (S3.31) shows that the $\boldsymbol{\Psi}_t(t)$ constructed by the $\boldsymbol{\Psi}(\pm\omega)$ given by Eqs. (S3.25)-(S3.27) consistently takes real values, which then ensures the legitimacy of using the PML-RQNMs of frequency-nondispersive PMLs for constructing the $\boldsymbol{\Psi}_t(t)$.

*The proof ends*.

In the following, we will provide a proof of Eq. (S3.30).

*Proof*: From Eq. (S3.27), one can obtain,

$$
\begin{aligned}
(F_m^+)^* &= \iiint_{\Omega_{\text{physical}}\cup\Omega_{\text{PML}}} \left\{ \mathbf{E}^+(\omega_m^+)^* \cdot \left\{ \left. \frac{\partial[\omega\boldsymbol{\varepsilon}^+(\omega)]}{\partial\omega} \right|_{\omega=\omega_m^+} \right\}^* \cdot \mathbf{E}^+(\omega_m^+)^* \right. \\
&\qquad \left. -\mathbf{H}^+(\omega_m^+)^* \cdot \left\{ \left. \frac{\partial[\omega\boldsymbol{\mu}^+(\omega)]}{\partial\omega} \right|_{\omega=\omega_m^+} \right\}^* \cdot \mathbf{H}^+(\omega_m^+)^* \right\} d^3\mathbf{r} \\
&= \iiint_{\Omega_{\text{physical}}\cup\Omega_{\text{PML}}} \left\{ \mathbf{E}^-(\omega_m^-) \cdot \left. \frac{\partial[\omega\boldsymbol{\varepsilon}^-(\omega)]}{\partial\omega} \right|_{\omega=\omega_m^-} \cdot \mathbf{E}^-(\omega_m^-) \right. \\
&\qquad \left. -\mathbf{H}^-(\omega_m^-) \cdot \left. \frac{\partial[\omega\boldsymbol{\mu}^-(\omega)]}{\partial\omega} \right|_{\omega=\omega_m^-} \cdot \mathbf{H}^-(\omega_m^-) \right\} d^3\mathbf{r} \\
&= F_m^- .
\end{aligned} \tag{S3.32}
$$

The second equality of Eq. (S3.32) is obtained by using Eq. (S3.24) and,

$$
\left\{ \left. \frac{\partial[\omega\boldsymbol{\varepsilon}^+(\omega)]}{\partial\omega} \right|_{\omega=\omega_m^+} \right\}^* = \left. \frac{\partial[\omega\boldsymbol{\varepsilon}^-(\omega)]}{\partial\omega} \right|_{\omega=\omega_m^-} \ \text{(a)},\quad \left\{ \left. \frac{\partial[\omega\boldsymbol{\mu}^+(\omega)]}{\partial\omega} \right|_{\omega=\omega_m^+} \right\}^* = \left. \frac{\partial[\omega\boldsymbol{\mu}^-(\omega)]}{\partial\omega} \right|_{\omega=\omega_m^-} \ \text{(b)}, \tag{S3.33}
$$

which will be proved in the following. To prove Eq. (S3.33a), under the assumption that the $\boldsymbol{\varepsilon}^{\pm}(\omega)$ is analytic at $\omega=\omega_m^{\pm}$, $\boldsymbol{\varepsilon}^{\pm}(\omega)$ can be expressed as Laurent series in a neighborhood of $\omega=\omega_m^{\pm}$,

$$
\boldsymbol{\varepsilon}^{\pm}(\omega) = \sum_{n=0}^{\infty} c_n^{\pm}(\omega-\omega_m^{\pm})^n,\ \omega\in\mathbb{C}. \tag{S3.34}
$$

From Eq. (S3.34), one can obtain,

$$
\begin{aligned}
\boldsymbol{\varepsilon}^+(\omega)^* &= \sum_{n=0}^{\infty}(c_n^+)^*[\omega^* - (\omega_m^+)^*]^n = \sum_{n=0}^{\infty}(c_n^+)^*(\omega^* + \omega_m^-)^n \\
&= \boldsymbol{\varepsilon}^-(-\omega^*) = \sum_{n=0}^{\infty} c_n^-(-\omega^* - \omega_m^-)^n = \sum_{n=0}^{\infty} c_n^-(-1)^n(\omega^* + \omega_m^-)^n,
\end{aligned} \tag{S3.35}
$$

where the second equality is obtained by using Eq. (S3.23a), and the third equality is obtained by using the first one of Eq. (S3.16b). Equation (S3.35) then yields,

$$
(c_n^+)^* = (-1)^n c_n^-,\ n=0,\ 1,\ 2,\ \cdots \tag{S3.36}
$$

From Eq. (S3.34), one can obtain,

$$
\left[ \left. \frac{\partial\boldsymbol{\varepsilon}^+(\omega)}{\partial\omega} \right|_{\omega=\omega_m^+} \right]^* = (c_1^+)^* = -c_1^- = -\left. \frac{\partial\boldsymbol{\varepsilon}^-(\omega)}{\partial\omega} \right|_{\omega=\omega_m^-}, \tag{S3.37}
$$

where the second equality is obtained from Eq. (S3.36). Then there is,

$$
\begin{aligned}
\left\{ \left. \frac{\partial[\omega\boldsymbol{\varepsilon}^+(\omega)]}{\partial\omega} \right|_{\omega=\omega_m^+} \right\}^* &= \left\{ \boldsymbol{\varepsilon}^+(\omega_m^+) + \omega_m^+ \left. \frac{\partial[\boldsymbol{\varepsilon}^+(\omega)]}{\partial\omega} \right|_{\omega=\omega_m^+} \right\}^* = \boldsymbol{\varepsilon}^+(\omega_m^+)^* + (\omega_m^+)^* \left\{ \left. \frac{\partial[\boldsymbol{\varepsilon}^+(\omega)]}{\partial\omega} \right|_{\omega=\omega_m^+} \right\}^* \\
= \boldsymbol{\varepsilon}^-(-(\omega_m^+)^*) + (\omega_m^+)^* \left\{ \left. \frac{\partial[\boldsymbol{\varepsilon}^+(\omega)]}{\partial\omega} \right|_{\omega=\omega_m^+} \right\}^* &= \boldsymbol{\varepsilon}^-(\omega_m^-) + \omega_m^- \left. \frac{\partial\boldsymbol{\varepsilon}^-(\omega)}{\partial\omega} \right|_{\omega=\omega_m^-} = \left. \frac{\partial[\omega\boldsymbol{\varepsilon}^-(\omega)]}{\partial\omega} \right|_{\omega=\omega_m^-},
\end{aligned} \tag{S3.38}
$$

where the 3rd equality is obtained from the first one of Eq. (S3.16b), the 4th equality is obtained from Eq. (S3.37) and Eq. (S3.23a), and the last equality simply gives Eq. (S3.33a). In a fully parallel way, Eq. (S3.33b) can also be proved.

*The proof ends.*

**D. Central symmetry of eigenfrequencies of PML-RQNMs for frequency-nondispersive PMLs and medium**

As shown in the inset of Fig. S1(a), the PML-RQNMs$^+$ or PML-RQNMs$^-$ of the single slab satisfy a central symmetry of eigenfrequencies with respect to the origin. This spectral central symmetry is due to the fact that the medium and the PMLs for the single slab are both frequency-nondispersive, as to be explained in the following.

For the case that the medium and the PMLs are both frequency-nondispersive, the source-free Maxwell's equations (S3.21a) satisfied by a PML-RQNM$^+$ with eigenfrequency $\omega_m^+$ and electromagnetic field $\mathbf{\Psi}'^+(\omega_m^+) = [\mathbf{E}'^+(\omega_m^+), \mathbf{H}'^+(\omega_m^+)]^\mathrm{T}$ become,

$$\nabla' \times \mathbf{E}'^+(\omega_m^+) = i\omega_m^+ \boldsymbol{\mu}'^+ \cdot \mathbf{H}'^+(\omega_m^+), \quad \nabla' \times \mathbf{H}'^+(\omega_m^+) = -i\omega_m^+ \boldsymbol{\varepsilon}'^+ \cdot \mathbf{E}'^+(\omega_m^+), \tag{S3.39}$$

where

$$\boldsymbol{\varepsilon}'^+(\omega) = \boldsymbol{\varepsilon}'^+, \ \boldsymbol{\mu}'^+(\omega) = \boldsymbol{\mu}'^+, \tag{S3.40}$$

are both independent of the complex-valued frequency $\omega$. Equation (S3.39) leads to,

$$\nabla' \times \mathbf{E}'^+(\omega_m^+) = i(-\omega_m^+)\boldsymbol{\mu}'^+ \cdot [-\mathbf{H}'^+(\omega_m^+)], \quad \nabla' \times [-\mathbf{H}'^+(\omega_m^+)] = -i(-\omega_m^+)\boldsymbol{\varepsilon}'^+ \cdot \mathbf{E}'^+(\omega_m^+), \tag{S3.41}$$

which implies that there must exist another PML-RQNM$^+$ with eigenfrequency $-\omega_m^+$ and electromagnetic field $\mathbf{\Psi}'^+(-\omega_m^+) = [\mathbf{E}'^+(-\omega_m^+), \mathbf{H}'^+(-\omega_m^+)]^\mathrm{T}$ and satisfying,

$$\mathbf{E}'^+(-\omega_m^+) = \mathbf{E}'^+(\omega_m^+), \ \mathbf{H}'^+(-\omega_m^+) = -\mathbf{H}'^+(\omega_m^+). \tag{S3.42}$$

For deriving Eq. (S3.42), we have used,

$$\boldsymbol{\varepsilon}'^+(\omega_m^+) = \boldsymbol{\varepsilon}'^+(-\omega_m^+) = \boldsymbol{\varepsilon}'^+, \ \boldsymbol{\mu}'^+(\omega_m^+) = \boldsymbol{\mu}'^+(-\omega_m^+) = \boldsymbol{\mu}'^+, \tag{S3.43}$$

which is an inference from Eq. (S3.40). Equation (S3.42) can lead to,

$$\mathbf{E}^+(-\omega_m^+) = \mathbf{L}^+ \cdot \mathbf{E}'^+(-\omega_m^+) = \mathbf{L}^+ \cdot \mathbf{E}'^+(\omega_m^+) = \mathbf{E}^+(\omega_m^+), \tag{S3.44a}$$

$$\mathbf{H}^+(-\omega_m^+) = \mathbf{L}^+ \cdot \mathbf{H}'^+(-\omega_m^+) = \mathbf{L}^+ \cdot [-\mathbf{H}'^+(\omega_m^+)] = -\mathbf{H}^+(\omega_m^+), \tag{S3.44b}$$

where Eqs. (S3.18b) and (S3.18c) are used. Equation (S3.42) or (S3.44) is just the spectral central symmetry satisfied by the PML-RQNMs$^+$ for frequency-nondispersive medium and PMLs. The spectral central symmetry of the PML-RQNMs$^-$ can be proven in the same way.

One may notice that the spectral central symmetry of Eq. (S3.44) is not satisfied by the physical QNMs of the single slab as shown by Eq. (29) in the main text, even though the medium is frequency-nondispersive. This is due to the outgoing-wave condition satisfied by the physical QNMs. Or more specifically, if one makes the assumption that there exist two physical QNMs satisfying Eq. (S3.44) with eigenfrequencies $\omega_m^+$ and $-\omega_m^+$, then one can infer that the two physical QNMs must have electromagnetic fields which have $z$-dependences $\exp(i\omega_m^+ n_1 z/c)$ and $\exp[i(-\omega_m^+)n_1 z/c]$ for $z>d$ and which simultaneously satisfy the outgoing-wave condition as $z\rightarrow+\infty$, which however is impossible, so the assumption is false.

**S4. Influence of the PML parameters on the accuracy of the QCT using PML-RQNMs**

For the complex coordinate transformation [Eq. (33) in the main text] that we used to implement the PML, a larger imaginary part of $f_{\rm PML}$ can achieve a stronger attenuation of electromagnetic field in the PML and resultantly a weaker electromagnetic field at the outer boundary of PML, which thus is expected to achieve a higher accuracy in solving the PML-RQNMs with the PWEM, and resultantly, a higher accuracy of the QCT using PML-RQNMs.

For a numerical test of the above expectation, Fig. S4(a) shows that with the increase of Im($f_{\rm PML}$) (all the other PML parameters being unchanged and consistent with those in the main text), the relative error $\delta$ (see its definition in Sec. III B2) of $\tilde{\omega}_{-,1}$ decreases significantly, which holds as well for the relative error $\sigma$ (see its definition in Sec. III B3) of scattered field at the excitation frequency $\omega=\mathrm{Re}(\tilde{\omega}_{-,1})$ as shown in Fig. S4(b).

For comparison, Fig. S4 also provides the corresponding results of the QCT using ESC-RQNMs, showing that the relative errors $\delta$ and $\sigma$ both decrease monotonically with the increase of the mode number $M$ of ESC-RQNMs, which is similar to the results corresponding to $\tilde{\omega}_{+,1}$ in the main text [Fig. 4(b)].

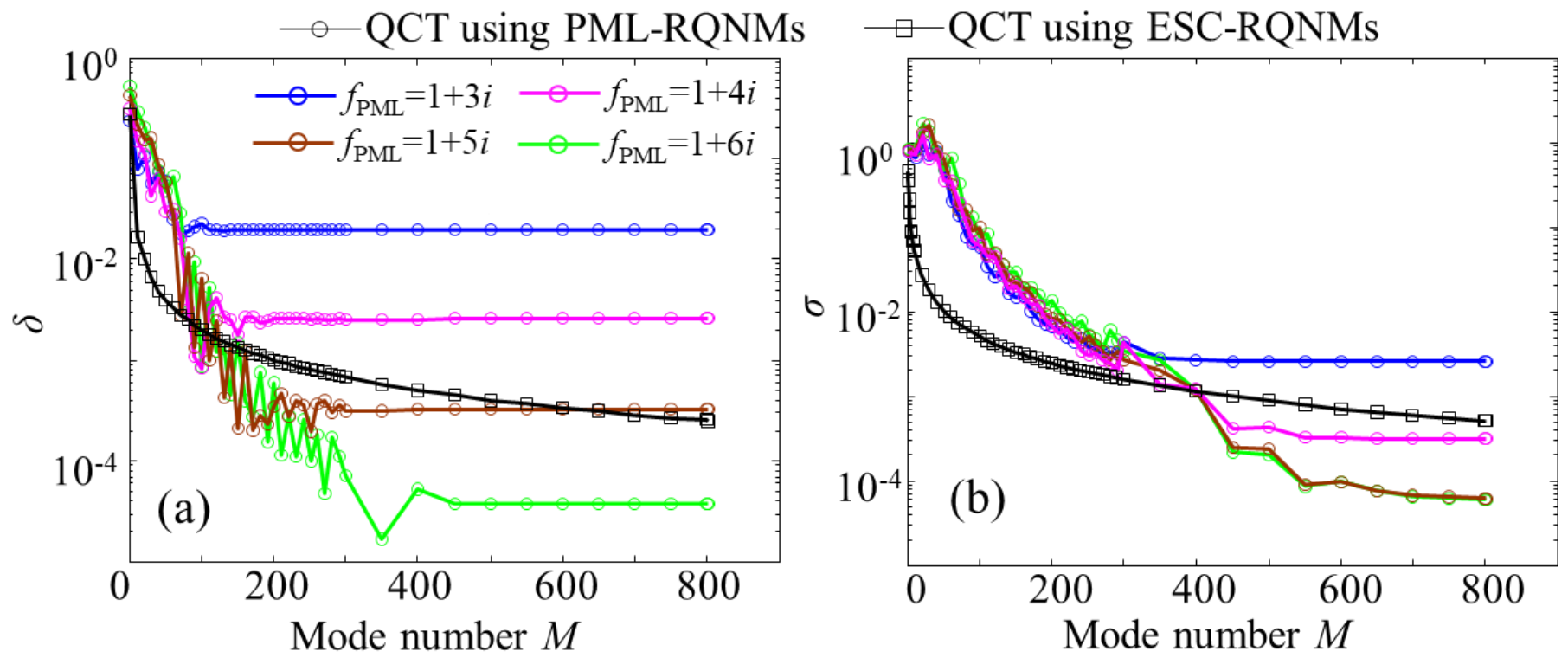

Fig. S4 (a) Convergence curves of the QCT in predicting the eigenfrequency $\tilde{\omega}_{-,1}$, where the relative error $\delta$ is plotted as a function of the mode number $M$. (b) Convergence curves of the QCT in predicting the scattered field excited by a plane wave at frequency $\omega=\mathrm{Re}(\tilde{\omega}_{-,1})$, where the relative error $\sigma$ is plotted as a function of $M$. In (a) and (b), the black squares show the results obtained with the QCT using ESC-RQNMs. The circles of different colors show the results of the QCT using PML-RQNMs with different values of Im($f_{\rm PML}$).

**S5. Results of the QCT using PML-RQNM$^-$ of a single slab in predicting the PML-RQNM$^-$ of the two coupled slabs**

Figure 4(a) in the main text shows that the physical QNMs (red dots) with a negative real part of eigenfrequency cannot be predicted by the QCT using PML-RQNMs. The reason is that only the PML-RQNMs$^+$ [see Eq. (S3.21) for the relevant definitions] of a single slab are used in the QCT, so that the QCT can only predict the PML-RQNMs$^+$ of the coupled slabs, which do not include the physical QNMs with a negative real part of eigenfrequency. In this section, we will show that if the PML-RQNMs$^-$ of a single slab are used in the QCT, then the QCT can predict the PML-RQNMs$^-$ of the coupled slabs, which will include the physical QNMs with a negative real part of eigenfrequency.

For solving the PML-RQNMs$^-$ of the single slab, although Eq. (S3.23) shows that they can be obtained from the PML-RQNMs$^+$ already solved in the main text, here they are solved by using the full-wave PWEM (with the same parameters as those in Fig. S3) so as to provide a complete numerical test of the QCT using PML-RQNMs$^-$.

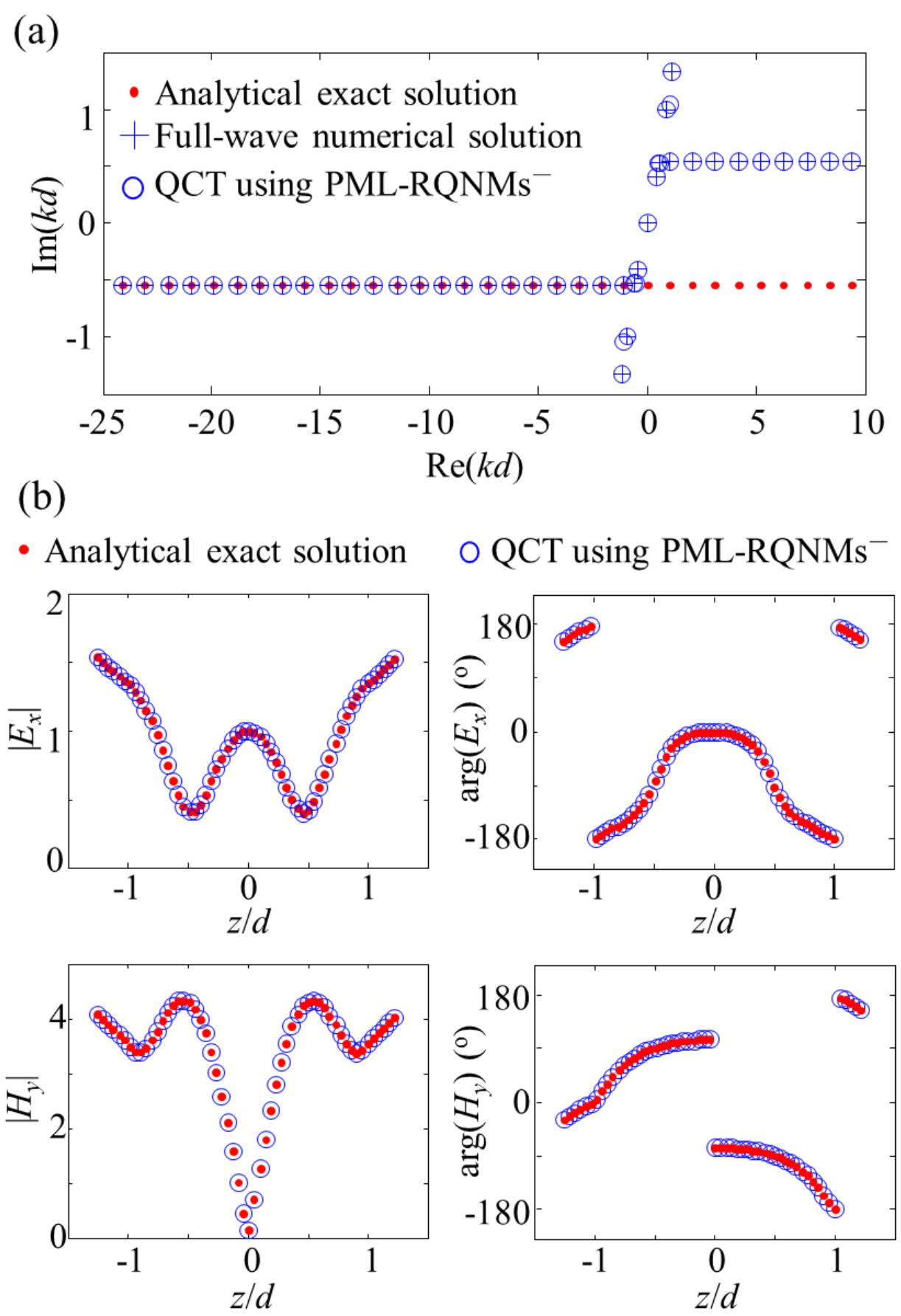

Fig. S5 (a) Eigenfrequencies of PML-RQNMs$^-$ of the two coupled slabs obtained with the full-wave PWEM (blue pluses) and the QCT using PML-RQNMs$^-$ of a single slab (blue circles). The red dots are the results of exact solution (i.e., physical QNMs). (b) Electromagnetic field for the PML-RQNM$^-$ of the two coupled slabs obtained with the QCT using PML-RQNMs$^-$ of a single slab (blue circles) and for the corresponding symmetric physical QNM of order $m$=−1 (with eigenfrequency $\tilde{\omega}_{+,-1}$, red dots).

Figure S5(a) shows the eigenfrequencies of PML-RQNMs$^-$ of the two coupled slabs. The predictions of QCT using PML-RQNMs$^-$ of a single slab (blue circles) agree well with the full-wave PWEM results (blue pluses, truncated harmonic order $M_z$=400). Here for solving the PML-RQNMs$^-$ of the coupled slabs by using the PWEM, we set $f_{\rm PML}$=1−3$i$, with all the other PML parameters consistent with those in the PWEM for solving the PML-RQNMs$^+$ of the coupled slabs [as shown in Fig. 4(a) of the main text where $f_{\rm PML}$=1+3$i$].

Notably, Fig. S5(a) shows that the physical QNMs [red dots, given by Eq. (29) in the main text with $d$ replaced by 2$d$] with a negative real part of eigenfrequency can be well predicted by some PML-RQNMs$^-$, which are thus classified into the physical PML-RQNMs (see the relevant discussions in Sec. II C of the main text).

Figure S5(b) shows the electromagnetic field for the symmetric physical QNM of order $m$=−1 [with eigenfrequency $\tilde{\omega}_{+,-1}$ specified after Eq. (30) in the main text]. Again, the analytical exact solution of physical QNM (red dots) can be well predicted by the QCT using PML-RQNMs$^-$ of a single slab (blue circles).

**References**

[1] E. A. Muljarov and T. Weiss, Resonant-state expansion for open optical systems: Generalization to magnetic, chiral, and bi-anisotropic materials, Opt. Lett. **43**, 1978 (2018).

[2] P. T. Kristensen, K. Herrmann, F. Intravaia, and K. Busch, Modeling electromagnetic resonators using quasinormal modes, Adv. Opt. Photonics **12**, 612 (2020).

[3] P. Lalanne, Effective properties and band structures of lamellar subwavelength crystals: Plane-wave method revisited, Phys. Rev. B **58**, 9801 (1998).

[4] J. P. Hugonin and P. Lalanne, Perfectly matched layers as nonlinear coordinate transforms: A generalized formalization, J. Opt. Soc. Am. A **22**, 1844 (2005).

[5] C. Sauvan, T. Wu, R. Zarouf, E. A. Muljarov, and P. Lalanne, Normalization, orthogonality, and completeness of quasinormal modes of open systems: The case of electromagnetism [Invited], Opt. Express **30**, 6846 (2022).

[6] C. Sauvan, J. P. Hugonin, I. S. Maksymov, and P. Lalanne, Theory of the spontaneous optical emission of nanosize photonic and plasmon resonators, Phys. Rev. Lett. **110**, 237401 (2013).

[7] L. Li, Use of Fourier series in the analysis of discontinuous periodic structures, J. Opt. Soc. Am. A **13**, 1870 (1996).

[8] W. Yan, R. Faggiani, and P. Lalanne, Rigorous modal analysis of plasmonic nanoresonators, Phys. Rev. B **97**, 205422 (2018).

[9] A. Gras, P. Lalanne, and M. Durufle, Nonuniqueness of the quasinormal mode expansion of electromagnetic Lorentz dispersive materials, J. Opt. Soc. Am. A **37**, 1219 (2020).

[10] G. Demésy, T. Wu, Y. Brûlé, F. Zolla, A. Nicolet, P. Lalanne, and B. Gralak, Dispersive perfectly matched layers and high-order absorbing boundary conditions for electromagnetic quasinormal modes, J. Opt. Soc. Am. A **40**, 1947 (2023).